# A Generalization of Relative Entropy to Count Vectors and its Concentration Property

BY KOSTAS N. OIKONOMOU*

*Email:* `ko56@winlab.rutgers.edu`

*April 2024*

**Abstract**

We introduce a new generalization of relative entropy to non-negative vectors with sums $>1$. We show in a purely combinatorial setting, with no probabilistic considerations, that in the presence of linear constraints defining a convex polytope, a concentration phenomenon arises for this generalized relative entropy, and we quantify the concentration precisely. We also present a probabilistic formulation, and extend the concentration results to it. In addition, we provide a number of simplifications and improvements to our previous work, notably in dualizing the optimization problem, in the concentration with respect to $\ell_\infty$ distance, and in the relationship to generalized KL-divergence. A number of our results apply to general compact convex sets, not necessarily polyhedral.

# Table of contents

---

*. Visiting Research Scientist, Rutgers University Winlab, North Brunswick NJ, 08902 USA.

# 1 Introduction

In previous work, [OG16], we studied the concentration of Shannon entropy for probability vectors, and in [Oik17] the concentration of a generalized entropy for count vectors, i.e. vectors whose elements are natural numbers. In this paper we extend the work of [Oik17] to the concentration of a generalized *relative entropy* for count vectors. Roughly speaking, when the concentration phenomenon can be established, it shows that the solution to an under-specified problem found by optimizing some entropy measure is the "most likely to be realized" by a large factor. Some may find a result of this kind more satisfying than other arguments, including axiomatic justifications of entropic inference.

    The logical strength of the concentration property is that it follows from a *purely combinatorial* argument, with minimal assumptions. Notably, it need not involve any notion of probability or randomness. This notion of concentration has similarities, but is not to be confused with, the probabilistic notion of concentration of measure[1.1]. [OG16] and [Oik17] provide more background to our combinatorial approach to entropy concentration, and Example 2.1 below gives a concise overview.

---

[1.1]. But see below about §7.

The generalized relative entropy $G(x\|y)$ presented here is an extension of the generalized entropy $G(x)$ of [Oik17]. Since $G(x\|y)$ is a new function, §2 is devoted to it. We first define some terminology and notation, and then introduce a two-dimensional extension of the combinatorial setting used in [Oik17] to interpret $G(x\|y)$. Next we establish a number of the function's basic properties such as non-negativity, concavity, etc. Finally, we look at its maximization subject to linear constraints, including the dualization of the optimization problem. In §3 we explore the relationship between $G(x\|y)$ and the commonly-used I-divergence (generalized KL-divergence), and show a new correspondence between optimizing these two measures. §4 and §5 are devoted to the concentration property of the generalized relative entropy: around its maximum value, and around the vector which maximizes it, respectively. §6 examines how the concentration increases as the problem becomes larger in a certain way which we call 'scaling'. Our main results in these sections are of two kinds: Lemmas 4.2 and 5.4 which apply to problems of a given, fixed size, and Theorem 6.1 which shows that when the problem size is increased by scaling, either type of concentration increases exponentially with the scaling factor. Lastly, in §7 we show that the purely combinatorial formulation of §4 and §5 easily extends to a probabilistic one[1,2]: Theorem 7.1 shows that the count vector with maximum generalized relative entropy under the specified prior (a probability vector) has exponentially greater *probability* than the set of all other count vectors far from it, either in entropy value or in distance. Some numerical examples illustrating the major results of the paper are given in §8.

The generalized relative entropy presented here contains the generalized entropy of [Oik17] as a simple special case, so we point out some important new or improved results. We give a much simpler definition of the optimal count vector $v^*$, and explore new aspects of the relationship with I-divergence. We formulate, and solve, the dual of the problem of maximizing $G(x\|y)$. We give a much tighter, and more intuitively-satisfying definition of the sets of vectors that are close to or far from in $\ell_\infty$ distance to the vector which maximizes $G(x\|y)$. Finally, our treatment of concentration with scaling unifies both the 'value' and 'distance' cases in a single result, Theorems 6.1 and 7.1, for the combinatorial and probabilistic formulations respectively.

## 2 The generalized relative entropy

Here we introduce the generalized relative entropy $G(x\|y)$ between two real non-negative vectors and present its basic properties. First we clarify some terminology and explain some notation.

**Terminology.** The terms relative entropy, cross-entropy, divergence, and KL-divergence have all been used to refer to the function $\sum_i x_i \ln \frac{x_i}{y_i}$. This function has been variously denoted $D(x\|y)$, $H(x\|y)$, $D_{\text{KL}}(x\|y)$, or $\text{KL}(x\|y)$. And sometimes "cross-entropy" is used for the function $H(x, y) = -\sum_i x_i \ln y_i$. For our generalized relative entropy we use the innovative notation $G(x\|y)$.

When we want to avoid probability connotations, we refer to what is usually called a probability vector as a "density vector", i.e. simply normalized to 1. Here we deal extensively with non-density or un-normalized vectors: sets of non-negative vectors, all of the same dimension $m \geqslant 2$, for which there is *no single constant* $s > 0$ such that all vectors in the set sum to $s$. E.g. with $m = 3$, both $(2, 0, 4)$ and $(1, 7.3, 17)$ can belong to such a set. If the elements of a non-density vector are natural numbers we refer to it as a 'count' vector, and if they are rational numbers with the same denominator we call it a 'frequency' vector.

---

1.2. For those who feel that simple combinatorial arguments leave something to be desired.

**Notation.** When $x$ is a vector, which should be clear from the context, we use $\geqslant$ etc. in the element-wise sense; likewise, if $a$ is a scalar, $ax$ denotes element-wise multiplication, as usual. $x, y, z, u, v$ will denote vectors in $\mathbb{R}_+^m$, whereas $\mu, \nu$ will denote vectors in $\mathbb{N}^m$. Density vectors corresponding to $x, y$, i.e. $x / \sum_i x_i$, $y / \sum_i y_i$, are denoted $\chi, \psi$.

## 2.1 Combinatorial setting, count vectors

Consider the process of allocating $n$ identical balls, one-by-one, to the $r$ bins shown in Fig. 2.1. The "balls and bins" paradigm is well-known in discussions of the combinatorial aspects of Shannon entropy, and we have used it in previous papers [Oik17], [OG16]. The array of bins is normally taken to be one-dimensional, but the idea of bins of varying sizes appeared in [SS06] in connection with relative entropy. Here we elaborate on this by making the array explicitly two-dimensional.

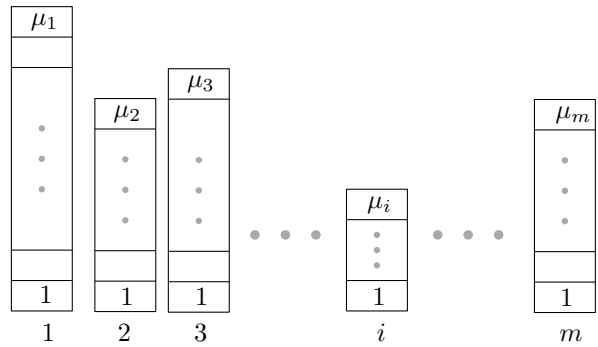

**Figure 2.1.** A 2-dimensional array of $r$ bins arranged in $m$ columns of sizes $\mu_1, ..., \mu_m$, $\sum_i \mu_i = r$. The bins are identical and have unlimited capacity. The shape of the array represents a 'prior' $\mu = (\mu_1, ..., \mu_m)$.

The balls are indistinguishable, but the bins are distinguishable, numbered 1 to $r$. So an 'allocation' is a sequence $\sigma = (\sigma_1, ..., \sigma_n)$ with $\sigma_q \in \{1, ..., r\}$; $\sigma_q = j$ means that the $q$th ball goes into bin $j$. The number of all possible allocations is $r^n$. The number of allocations such that $\nu_1$ balls end up in the bins of column 1, $\nu_2$ in the bins of column 2, ..., and $\nu_m$ in those of column $m$ is

$$\#_\mu \nu = \binom{\nu_1 + \cdots + \nu_m}{\nu_1, ..., \nu_m} \mu_1^{\nu_1} \cdots \mu_m^{\nu_m}, \tag{2.1}$$

where $\nu = (\nu_1, ..., \nu_m)$ is the vector of column occupancies. We will also refer to $\nu$ as the "column sum count vector", and to the sequences $\sigma$ that give rise to it as its 'realizations'. The vector $\mu$ specifies the *shape* of the array of bins; if we think in terms of Bayesian inference, we may view it as a prior. There are two other ways of viewing the allocation process that are helpful in understanding the role of the prior $\mu$:

- We have $r$ distinguishable objects, the bins, which are of $m$ kinds, $\mu_1$ of kind 1, $\mu_2$ of kind 2, ..., with $\sum_i \mu_i = r$. In how many ways can we choose $\nu_1$ objects of kind 1, $\nu_2$ of kind 2, ..., and $\nu_m$ of kind $m$? (A particular object can be chosen many times.)
- There are $m$ bins (not $r$), each of which is divided into sub-bins, $\mu_i$ for bin $i$. First $\nu_1, ..., \nu_m$ balls are allocated to the bins, i.e. the columns of Fig. 2.1, and then each $\nu_i$ is further allocated to the sub-bins of bin $i$.

The following very simple example, adapted from [Oik17], clarifies some points that are fundamental to the rest of the paper.

**Example 2.1.** Indistinguishable balls are to be placed one-by-one in a 3-column array of bins, colored red, green, and blue respectively. The final totals $(\nu_r, \nu_g, \nu_b)$ of the *columns* must satisfy $\nu_r + \nu_g = 4$ and $\nu_g + \nu_b \leqslant 6$. *Any* number of balls satisfying these constraints can be used. If each column consists of just one bin, i.e. $\mu = (1,1,1)$, the top part of Table 2.1 lists all the count vectors that satisfy the constraints, their sums $n$, and their *number of realizations* $\#_\mu \nu$ given by (2.1). What can be said about the "most likely" final content of the bins, i.e. the vector $\nu^*(\mu)$ with largest $\#_\mu \nu$? And what if the bin array has columns of sizes $\mu^{(2)} = (3,3,3)$ or $\mu^{(3)} = (1,3,1)$?

| $\nu_r$ | $\nu_g$ | $\nu_b$ | $n$ | $\#_{\mu^{(1)}}\nu$ |
|---|---|---|---|---|
| 0 | 4 | 0 | 4 | 1 |
| 4 | 0 | 0 |   | 1 |
| 1 | 3 | 0 |   | 4 |
| 2 | 2 | 0 |   | 6 |
| 3 | 1 | 0 |   | 4 |
| 3 | 0 | 1 |   | 4 |
| 0 | 4 | 1 | 5 | 5 |
| 1 | 3 | 1 |   | 20 |
| 2 | 2 | 1 |   | 30 |
| 3 | 1 | 1 |   | 20 |
| 4 | 0 | 1 |   | 5 |

| $\nu_r$ | $\nu_g$ | $\nu_b$ | $n$ | $\#_{\mu^{(1)}}\nu$ |
|---|---|---|---|---|
| 0 | 4 | 2 | 6 | 15 |
| 1 | 3 | 2 |   | 60 |
| 2 | 2 | 2 |   | 90 |
| 3 | 1 | 2 |   | 60 |
| 4 | 0 | 2 |   | 15 |
| 1 | 3 | 3 | 7 | 140 |
| 2 | 2 | 3 |   | 210 |
| 3 | 1 | 3 |   | 140 |
| 4 | 0 | 3 |   | 35 |

| $\nu_r$ | $\nu_g$ | $\nu_b$ | $n$ | $\#_{\mu^{(1)}}\nu$ |
|---|---|---|---|---|
| 2 | 2 | 4 | 8 | 420 |
| 3 | 1 | 4 |   | 280 |
| 4 | 0 | 4 |   | 70 |
| **3** | **1** | **5** | **9** | **504** |
| 4 | 0 | 5 |   | 126 |
| 4 | 0 | 6 | 10 | 210 |

|  | $\nu_r^*$ | $\nu_g^*$ | $\nu_b^*$ | $\#_\mu \nu$ |
|---|---|---|---|---|
| $\mu^{(1)}$ | 3 | 1 | 5 | 504 |
| $\mu^{(2)}$ | 4 | 0 | 6 | 12400290 |
| $\mu^{(3)}$ | 1 | 3 | 3 | 3780 |
|  | 2 | 2 | 4 | 3780 |

**Table 2.1.** TOP: the column sum count vectors $\nu = (\nu_r, \nu_g, \nu_b)$ satisfying $\nu_r + \nu_g = 4, \nu_g + \nu_b \leqslant 6$, their sum $n$, and their number of realizations $\#_\mu \nu$ under the prior $\mu^{(1)} = (1,1,1)$. BOTTOM: the optimal count vectors under $\mu^{(1)}$, $\mu^{(2)} = (3,3,3), \mu^{(3)} = (1,3,1)$.

The example makes three fundamental points. First, it is not possible to find a *single frequency* vector that can be naturally associated with the problem; without that, one cannot think about maximizing the Shannon entropy, or minimizing the ordinary relative or relative entropy if the bin array is not 'flat'[2.1]. Second, the fact that we don't know exactly how many balls are allocated may not seem terribly exciting: using the *largest* possible number of balls, 10 in this case, should lead to the answer. But this is not so: $\nu^*(\mu^{(1)}) = (3,1,5)$ sums to 9, and even $\nu$ summing to 8 have more realizations than the $\nu$ summing to 10. Finally, the shape $\mu$ of the array, which we may think of as a prior in Bayesian terms, behaves as intuitively expected: $\mu^{(2)}$ shifts the inference to larger $n$, with $\nu^*(\mu^{(2)}) = (4,0,6)$, whereas $\mu^{(3)}$ biases it toward $\nu_g$, with $\nu^*(\mu^{(3)}) = (1,3,3)$ and $(2,2,4)$.

## 2.2 Definition and basic properties

The generalized entropy $G(x)$ of a real $m$-vector $x$, introduced in [Oik17], is

$$G(x) \triangleq -\sum_i x_i \ln x_i + \left(\sum_i x_i\right) \ln\left(\sum_i x_i\right) = \left(\sum_i x_i\right) H(\chi), \quad x \geqslant 0, \tag{2.2}$$

---

2.1. For example, in the usual MAXENT problem we have a single $n$, and the distinction between *count* and *frequency* vectors doesn't really matter, there is a 1-1 correspondence; but this is not true here.

where $H(\chi)$ is the Shannon entropy of the normalized or density version $\chi = x/\sum_i x_i$ of $x$. Here we are using the convention $0 \ln 0 \triangleq 0^{2.2}$.

For $m$-vectors $x, y$ we define the *generalized relative entropy* $G(x\|y)$ of $x$ with respect to $y$ as

$$G(x\|y) \triangleq -\sum_i x_i \ln \frac{x_i}{y_i} + \left(\sum_i x_i\right) \ln\left(\sum_i x_i\right), \qquad x \geq 0, y > 0. \tag{2.3}$$

[The first term can be thought of as $-D(x\|y)$, the relative entropy or divergence, formally extended to non-density vectors.] To interpret $G(x\|y)$, recall that the *probability* of a *frequency* vector $f = \nu/n$, $n = \sum_i \nu_i$, under a p.d. $q$ is approximately exponential in the relative entropy $D(f\|q)$

$$\Pr_q(f) \approx e^{-nD(f\|q)},$$

where the $\approx$ means within a factor polynomial in $n$ (see e.g. [CT06], Theorem 11.1.4). Likewise, the *number of realizations* of a column sum *count* vector $\nu$ with the array of bins of Fig. 2.1 of shape $\mu$ is

$$\#_\mu \nu \approx e^{G(\nu\|\mu)}.$$

Further, if we divide both sides of (2.1) by $r^n$ the number of realizations becomes the *probability* of the count vector $\nu$ under the (rational) p.d. $\mu/r$; this is explored further in §7. Finally, whereas we are usually interested in minimizing the ordinary relative entropy/divergence $D(\cdot\|\cdot)$, we will generally want to *maximize* the generalized relative entropy $G(\cdot\|\cdot)$.

The following proposition lists some basic properties of $G(x\|y)$. Its proof, as well as the proofs of all other formal statements, is in the Appendix.

**Proposition 2.2.**

1. *For $y \geq 1$, $G(x\|y) \geq 0$, and if at least one element of $x$ is non-zero then $G(x\|y) > 0$. For any $y > 0$, $G(\mathbf{0}\|y) \triangleq 0$.*

2. *When $y$ is a density vector, i.e. $\sum_i y_i = 1$, $G(x\|y) \leq 0$.*

3. *Special values: $G(x\|\mathbf{1}) = G(x)$, $G(y\|y) = (\sum_i y_i) \ln (\sum_i y_i)$.*

4. *When $y \geq 1$, $G(x\|y)$ is a monotone increasing function of $x$, i.e. $x' \geq x \Rightarrow G(x'\|y) \geq G(x\|y)$. If there is one strict inequality in the l.h.s., there is strict inequality in the r.h.s.*

5. *$G(x\|y)$ is positively homogeneous in $x$, i.e. $G(\mathbf{0}\|y) = 0$ and for any $\alpha > 0$, $G(\alpha x\|y) = \alpha G(x\|y)$.*

6. *$G(x\|y)$ is a concave function of $x$, but it is not strictly concave.*

7. *$G(x\|z)$ is superlinear in $x$, i.e. $\forall x, y \geq 0$, $\forall \alpha, \beta \geq 0$, $G(\alpha x + \beta y\|z) \geq \alpha G(x\|z) + \beta G(y\|z)$.*

8. *For any $y > 0$, $G(x\|y)$ is bounded as*

$$\left(\sum_i x_i\right) \ln \left(\min_i y_i\right) \leq G(x\|y) \leq \left(\sum_i x_i\right) \ln \left(\sum_i y_i\right).$$

9. *Grouping of $x$ and $y$ increases $G(x\|y)$, i.e. for any $i, j$,*

$$G\big((x_1, ..., x_i + x_j, ..., x_m) \| (y_1, ..., y_i + y_j, ..., y_m)\big) \geq G(x\|y).$$

---

2.2. And are being somewhat sloppy by defining $G(x)$ for all $x \geq 0$ instead of $x > 0$; our definition makes it not differentiable on the boundary.

A consequence of (1), (2), (4) is that when $y \geqslant 1$, $G(x\|y)$ increases without bound as $x$ increases; but when $y$ is a density vector, $G(x\|y)$ is bounded above by 0. These facts play a role in the maximization of $G(x\|y)$ in §2.3.

The positive homogeneity property (5) has many important consequences that will be seen in the sequel. One of them is that if we have a point $x$ at which $G$ has the value $g_1$ and we want to find a $z$ where it takes the value $g_2$, this is $z = \alpha x$ with $\alpha = g_2/g_1$, if $\alpha > 0$.

Property (7), so-called by analogy with sublinearity[2.3], is more general than concavity. For property (9), note that the multinomial expression (2.1) behaves likewise[2.4].

Now we note some analogies with ordinary (Shannon) entropy $H(\chi) = -\sum_i \chi_i \ln \chi_i$ and ordinary divergence or relative entropy $D(\chi\|\psi) = \sum_i \chi_i \ln(\chi_i/\psi_i)$. We use $x,y$ to denote general vectors and $\chi, \psi$ to denote the corresponding density or probability vectors:

$$\begin{aligned} -D(\chi\|\psi) &= H(\chi) + \sum_i \chi_i \ln \psi_i, \\ G(x\|y) &= G(x) + \sum_i x_i \ln y_i, \\ G(\chi\|\psi) &= -D(\chi\|\psi), \end{aligned} \quad (2.4)$$

The last line follows from the first and the fact that for density vectors $\chi$ the generalized entropy $G(\chi)$ of (2.2) coincides with the Shannon entropy $H(\chi)$. So for density vectors, *minimizing* the relative entropy/divergence $D(\cdot\|\cdot)$ corresponds to *maximizing* the generalized relative entropy $G(\cdot\|\cdot)$. [Two density vectors are equal iff the corresponding general vectors are proportional, i.e. $\psi = \chi \Leftrightarrow \exists \alpha > 0 : y = \alpha x$.]

**Remark 2.3.** Expressions involving $\chi, \psi$ together with $x, y$ such as (2.2), (2.4) above, or (2.5) below, are useful only in a context where the sums of $x$ and $y$ can be treated as known constants. Otherwise each $\chi_i$ is a function of all of $x_1, \ldots, x_m$, and similarly for $\psi_i$. E.g., if $i \neq j$ we have $\partial \chi_i / \partial x_j = x_i/(x_1 + \cdots + x_m)^2$.

There is also a relationship between the generalized relative entropy of two vectors and the relative entropy of their normalized (density) versions, as well as bounds involving the $\ell_1$ distance between the density vectors:

$$\begin{aligned} G(x\|y) &= s \ln t - s D(\chi\|\psi), \\ G(x\|y) &\leqslant s \ln t - \frac{s}{2} \|\chi - \psi\|_1^2, \\ G(x\|y) &\geqslant s \ln t - \frac{s}{\psi_{\min}} \|\chi - \psi\|_1^2, \end{aligned} \quad (2.5)$$

where $s \triangleq \sum_i x_i$, $t \triangleq \sum_i y_i$, and $\chi \triangleq x/s$, $\psi \triangleq y/t$, and $\psi_{\min}$ is the smallest element of $\psi$. [Multiplying both sides of the first line in (2.4) by $s$ and using the fact that $G(x) = s H(\chi)$, we get $-s D(\chi\|\psi) = G(x) + \sum_i x_i \ln \psi_i$. Substituting $\psi_i = y_i/t$ in the r.h.s., and using the 2nd line of (2.4) we obtain the first line of (2.5). The second line follows from Pinsker's inequality $D(\chi\|\psi) \geqslant \frac{1}{2}\|\chi - \psi\|_1^2$, and the third line from the "reverse Pinsker" inequality $D(\chi\|\psi) \leqslant 1/\psi_{\min} \|\chi - \psi\|_1^2$ appearing in eq. (10) of [Sas15].]

---

[2.3]. We call $G$ 'superlinear' instead of saying that $-G$ is sublinear. In either case we assume that $G(x\|z)$ is suitably extended to $\bar{\mathbb{R}}$.

[2.4]. However, recall that the Shannon entropy *decreases* under grouping ([CT06], Ch. 2). And so does the generalized entropy $G(x)$.

Finally, we can consider changing the prior. We then have

$$\begin{aligned}G(x\|y)-G(x\|z) &= \sum_i x_i \ln \frac{y_i}{z_i}, \\ G(x\|\alpha y) &= G(x\|y)+(\sum_i x_i)\ln\alpha, \quad \alpha>0.\end{aligned} \quad (2.6)$$

These follow from (2.4) and (2.6) respectively.

## 2.3 Maximization under linear constraints

Suppose we have linear equality and inequality constraints $Ax=b, Cx\leqslant d$, where $A,b,C,d$ are real matrices and vectors. We assume that these constraints define a convex bounded polyhedron (convex polytope) in $\mathbb{R}^m_+$. For fixed $y>0$, $G(x\|y)$ is a concave function of $x$, so the problem

$$\max_{x\in \mathcal{C}(0)} G(x\|y), \qquad \mathcal{C}(0) \triangleq \{x\in \mathbb{R}^m_+ | Ax=b, Cx\leqslant d\} \quad (2.7)$$

is is a concave program (i.e. concave objective, convex inequality constraints, linear equality constraints) and any local maximum is global. The Lagrangian is

$$L(x,\lambda,\xi) = G(x\|y) - \lambda\cdot(Ax-b) - \xi\cdot(Cx-d),$$

where · denotes the dot product[2.5]. Further, assuming all functions are differentiable[2.6], the KKT conditions

$$\begin{aligned} &Ax^*=b,\ Cx^*\leqslant d, \\ &x_j^* = y_j\left(\sum_i x_i^*\right)e^{-(\lambda^*\cdot A_{.j}+\xi^*\cdot C_{.j})}, \quad 1\leqslant j\leqslant m, \\ &\xi_k^*(C_{.k}\cdot x^* - d_k)=0, \quad \xi_k^*\geqslant 0, \end{aligned} \quad (2.8)$$

where $A_{.j}$ is the $j$th column of $A$ and $k$ indexes the inequality constraints, are necessary and sufficient for the triple $(x^*,\lambda^*,\xi^*)$ to solve (2.7), irrespective of concavity. [See e.g. [BV04], §5.5.3.].

The expression in the second line of (2.8) determines the density vector $\chi^*$ as a function of $y$ and of the optimal multipliers or dual variables $\lambda^*,\xi^*$. When $G^*$ is known, $\sum_i x_i^*$, hence the $x_i^*$, can be found from $\chi^*$ by using the first line of (2.5).

**Remark 2.4.** On zeros in the solution: if the multipliers are to be finite, the expression for $x_j^*$ above does not admit $x_j^* = 0$. To avoid introducing special cases for handling 0s in the sequel, we will assume that all elements of $x^*$ that are forced to 0 by the constraints are removed from the solution. (See Example 8.1, and Remark 2.1 in [Oik17] for more details on this issue.) A consequence of the assumption that $x^*>0$ is that by Proposition 2.2(1), we can assume that if $y\geqslant 1$, $G(x^*\|y)$ is strictly positive.

In the sequel we use the notation

$$G^*(y) = G(x^*\|y) \triangleq \max_{x\in\mathcal{C}(0)} G(x\|y), \quad y>0. \quad (2.9)$$

---

2.5. We normally denote the dot product of $x,y$ by $x^Ty$, but sometimes also by $x\cdot y$, especially when the $^T$ becomes cumbersome, as in (2.8) below.

2.6. There is a technicality here, already noted in connection with (2.2).

In view of Proposition 2.2(4), for a maximum to exist when $y \geqslant 1$ it is necessary that all elements of $x$ be bounded from above; this is equivalent to requiring an upper bound on the sum of $x$. (Again equivalently, the polyhedron $C(0)$ is bounded, i.e. it is a polytope.) It is easy to see that condition (2.8) is unsatisfiable if this is not so. *We will therefore assume* that when $y \geqslant 1$, which will be the case in §4, §5, and §6, the constraints on $x$ impose (finite) bounds $s_1, s_2$ on its sum, which can be found by solving linear programs:

$$s_1 \triangleq \min_{x \in C(0)} x_1 + \cdots + x_m, \qquad s_2 \triangleq \max_{x \in C(0)} x_1 + \cdots + x_m, \qquad s_1, s_2 > 0. \tag{2.10}$$

How do these bounds depend on $A, b, C, d$? See §3.3 of [Oik17].

By putting the second line of (2.8) in (2.3), and noting that $x^*$ satisfies the equality constraints and some of the inequality constraints with equality (the 'binding' or 'active' inequalities at $x^*$), we find that the maximum value of the generalized relative entropy over $C(0)$ can be expressed in terms of the optimal multipliers/dual variables $\lambda^*, \xi^*$ and the data $b, d$ as

$$G^*(y) = \lambda^* \cdot b + \xi^* \cdot d, \tag{2.11}$$

where we recall that the $\xi_j^*$ corresponding to non-binding/inactive inequalities are 0 [2.7]. This expression will re-appear in §2.4.

Below we list some basic facts about maximizing $G(x \| y)$ with respect to $x$. Concerning $y$, we restrict attention to the two cases for which we have applications: $y \geqslant 1$, and $y$ is a density vector.

**Proposition 2.5.**

1. *When $y \geqslant 1$, $G(x \| y)$ is unbounded unless every element of $x$ (equivalently its sum) is bounded above.*
2. *When $y$ is a density vector and there are no constraints, $G(x \| y)$ attains its maximum of 0 at any point $x$ such that $\chi = y$, equivalently $x = \alpha y$ for some $\alpha > 0$.*
3. *For any $y > 0$, the maximum of $G(x \| y)$ under just the constraint $\sum_i x_i \leqslant s$ or $\sum_i x_i = s$ is $G^*(y) = s \ln r$, and occurs at $x^* = (s/r) y$, where $r$ is the sum of $y$.*
4. *In general, where the constraints $Ax = b$ and $Cx \leqslant d$ imply $s_1 \leqslant \sum_i x_i \leqslant s_2$, we have $G^*(y) \leqslant s_2 \ln r$. If $x^*$ is known, then $G^*(y) \leqslant s^* \ln r$.*

Part 1, which follows from Proposition 2.2(4), accords with the combinatorial interpretation of $G(\nu \| \mu)$. Parts 2 and 3 say that $G(\cdot \| \cdot)$ *shares a fundamental property of the divergence $D(\cdot \| \cdot)$*: when the prior is a density and there is no extra information, the 'posterior' is any vector with the same density; when the prior is $\geqslant 1$ and the posterior's sum is bounded above, the posterior is just a (scalar) multiple of the prior; further, the only dependence of the posterior on the prior is through the prior's *density*. These are fundamental, intuitive consistency properties of the posterior, or of the inference process. Finally, part 4 gives some easy bounds on $G^*(y)$.

Up to this point the issue of the *uniqueness* of the maximizer $x^*$ has been left open. In fact, as can be seen from part 2 of the proposition in some situations the maximizer is clearly not unique. Although $G(x \| y)$ is concave, recall Proposition 2.2(5), it is not strictly concave: e.g. consider $G((x_1, x_2) \| (1, 1))$. It is therefore not immediate that there is only one $x^*$ of the form in (2.8) that maximizes $G$. Nevertheless, this is almost always the case. The following result applies to $C(0)$, and to its generalization $C(\delta)$ introduced in §2.5, but is more generally valid, so we state it in its general form:

---

[2.7]. This is a property shared with the generalized entropy $G$ of [Oik17], as well as with the Shannon entropy $H$ (see [KK92]).

**Lemma 2.6.** *Let $C$ be a compact convex set. Then*

1. *If $y \geqslant 1$, $G(x\|y)$ is maximized at a unique point $x^* \in C$.*

2. *If $y$ is a density vector, the maximizer $x^*$ is unique if there is no $x \in C$ such that $G(x\|y) = 0$.*

3. *Further, in case 1, $x^*$ lies on the boundary of $C$[2.8].*

The uniqueness of $x^*$ is used in the proofs of some of the subsequent results. It is also useful when it comes time to adopt a specific optimal solution for a given problem.

Finally, suppose we scale by $\alpha > 0$ the *data* vectors $b, d$ appearing in the maximization problem (2.7). The fundamental property of this procedure is that *all aspects* of the solution scale by the same factor $\alpha$:

**Proposition 2.7.** *Let $x^*$ maximize $G(x\|y)$ under the linear constraints $Ax = b, Cx \leqslant d$, and let $\alpha > 0$ be any constant. Then the vector $\alpha x^*$ maximizes $G(x\|y)$ under the scaled constraints $Ax = \alpha b, Cx \leqslant \alpha d$, the maximum value of $G$ is $\alpha G(x^*\|y)$, and the new bounds on $\sum_i x_i$ are $\alpha s_1, \alpha s_2$.*

This result is basically due to the positive homogeneity of $G(x\|y)$ with respect to $x$, and will play a central role in §6.

## 2.4 Dualization

In general, dualization of an optimization problem may have computational advantages for its solution; here it is also useful for theoretical reasons, as will be seen in §5.

We refer to [RW09], [BV04] for terminology and background. In keeping with these references, our discussion of dualization is in terms of minimizing a function rather than of maximizing it[2.9]. So we will use

$$F(x\|\mu) \triangleq -G(x\|\mu) = \sum_i x_i \ln(x_i/\mu_i) - \left(\sum_i x_i\right) \ln\left(\sum_i x_i\right). \tag{2.12}$$

Our original, primal, problem $\max_{x \in C(0)} G(x\|\mu)$ is then equivalent to $\min_{x \in C(0)} F(x\|\mu)$ with $G^* = -F^*$. Here we show that the Lagrange dual of this is the problem

$$\max_{(\lambda,\xi) \in \mathcal{E}, \xi \geqslant 0} -b^T \lambda - d^T \xi,$$
$$\text{where} \quad \mathcal{E} = \{(\lambda,\xi) \in \mathbb{R}^l \mid \sum_i \mu_i e^{-(A^T\lambda + C^T\xi)_i} \leqslant 1\}, \tag{2.13}$$

with $l$ being the total number of constraints[2.10]. This is a concave problem, see (2.15) below.

We also show that there is no duality gap, i.e. if $\lambda^*, \xi^*$ solve (2.13) then

$$F^* \triangleq \min_{x \in C(0)} F(x\|\mu) = -b^T \lambda^* - d^T \xi^*. \tag{2.14}$$

---

2.8. But not necessarily on its *relative* boundary; see the proof.

2.9. Further, in this section we adopt the matrix-vector notation of [BV04], i.e. all vectors are column vectors and $^T$ denotes transpose. On the other hand, we use the convention of [RW09] that all functions are defined over the extended real numbers $\bar{\mathbb{R}}$, rather than specifying their effective domains as [BV04] do.

2.10. The different notations used here and in (2.8) can become confusing: note that $(A^T\lambda)_i = \lambda \cdot A_{\cdot i}$.

Modulo the difference in notation for the dot product, the negative of the r.h.s. above equals the r.h.s. of (2.11). This explains the provenance of (2.11): it is a consequence of dualization with no duality gap (strong duality).

Typically $m \gg l$, i.e. there will be many more variables than constraints in the primal problem, so when $m$ is large the dual (2.13) will have significantly fewer variables than the primal $\min_{x \in C(0)} F(x \| \mu)$ and should be easier to solve.

The solution $x^*$ of the primal problem (2.7) is unique, so it can be recovered from $\lambda^*, \xi^*$, and $F^* = -G^*$ by first using (2.8) to determine $\chi^*$ and then the first line of (2.5) to find $s^* = \sum_i x_i^*$. Also, by adding both sides of (2.8) over $j$ we see that the solution $\lambda^*, \xi^*$ of (2.13) satisfies the defining condition of the set $\mathcal{E}$ with equality.

We now turn to establishing (2.13). The Lagrangian for the problem $\min_{x \in C(0)} F(x \| \mu)$ is

$$L(x, \lambda, \xi) = F(x \| \mu) + \lambda^T (Ax - b) + \xi^T (Cx - d), \quad x \in \mathbb{R}^m$$

so the Lagrange dual function is

$$\begin{aligned}
\ell(\lambda, \xi) &\triangleq \inf_{x \in \mathbb{R}^m} L(x, \lambda, \xi) \\
&= \inf_{x \in \mathbb{R}^m} (F(x \| \mu) + (\lambda^T A + \xi^T C) x) - \lambda^T b - \xi^T d \\
\text{and by setting } y^T &= -(\lambda^T A + \xi^T C), \\
&= - \sup_{x \in \mathbb{R}^m} (y^T x - F(x \| \mu)) - \lambda^T b - \xi^T d \\
&= -F^\dagger(y \| \mu) - b^T \lambda - d^T \xi,
\end{aligned}$$

where $F^\dagger$ denotes the conjugate of $F$. Next, recalling that the indicator function of a set is 0 for elements in the set and $\infty$ otherwise,

**Proposition 2.8.** *The conjugate $F^\dagger(u \| \mu)$ of $F(x \| \mu)$ is the (convex) indicator function of the set $\{u \in \mathbb{R}^m \mid \sum_i \mu_i e^{u_i} \leqslant 1\}$.*

Consequently the dual of $L(x, \lambda, \xi)$ is the concave function

$$\ell(\lambda, \xi) = -\delta_{\mathcal{E}}(\lambda, \xi) - b^T \lambda - d^T \xi \tag{2.15}$$

where $\delta_{\mathcal{E}}$ is the indicator function of the set $\mathcal{E}$ in (2.13). The dual problem is to maximize this function, and this can be expressed as (2.13).

That there is strong duality, or 0 duality gap, i.e. (2.14), follows either from (2.8) with the differentiability assumption, or, without it, from the fact that the primal problem is convex, is assumed feasible, and involves only affine constraints, so Slater's constraint qualification is satisfied (see e.g. [BV04], §5.2.3).

## 2.5 The optimal count vector and tolerances on constraints

We introduce *tolerances* in the constraints appearing in (2.7) for three reasons. First, they are necessary to guarantee the existence of *integral* solutions (count vectors) to a problem whose data is arbitrary real numbers. Second, to ensure that the results in §6 on concentration in large problems hold *for all* scale factors larger than a certain "concentration threshold". Third, they may be used to reflect some uncertainty in the data. More details on this subject are in [Oik17] and [OG16].

We regard the matrices $A, C$ in (2.7) as specifying the *structure* of the constraints, and the vectors $b, d$ as specifying their *values*. Our tolerances ares only on the values, not on the structure of the constraints. Accordingly, a simple generalization of the set $C(0)$ in (2.7) is

$$C(\delta) \triangleq \{x \in \mathbb{R}_+^m \mid b - \delta \tilde{b} \leqslant Ax \leqslant b + \delta \tilde{b}, Cx \leqslant d + \delta \tilde{d}\}, \qquad \delta \geqslant 0, \tag{2.16}$$

where the 'error' vectors $\tilde{b}, \tilde{d}$ are $|b|, |d|$, except that any elements that are 0 are replaced by appropriate small positive constants. Although the definition does not require it, in any applications one would most likely want $\delta < 1$.

Given the solution $x^* \in C(0)$ of problem (2.7), we want to define an approximately optimal *integral* vector $\nu^* \in \mathbb{N}^m$. An obvious choice is to round $x^*$ element-wise to the nearest integer[2.11], and it can be seen that

**Proposition 2.9.** *The optimal count vector $\nu^* \triangleq [x^*]$ with $\sum_i \nu_i^* \triangleq n^*$ is such that*

$$|n^* - s^*| \leqslant \frac{m}{2}, \quad \|\nu^* - x^*\|_\infty \leqslant \frac{1}{2}, \quad \|\nu^* - x^*\|_1 \leqslant \frac{m}{2}, \quad \|f^* - \chi^*\|_1 \leqslant \frac{m}{n^*}.$$

The main point about tolerances is that for suitable $\delta$, the integral vector $[x]$ obtained by rounding any $x \in C(0)$ belongs to $C(\delta)$:

**Proposition 2.10.** ([Oik17]) *With $\tilde{b}, \tilde{d}$ as in (2.16), set*

$$\rho_\infty \triangleq \min(\tilde{b}_{\min} / \|\|A\|\|_\infty, \tilde{d}_{\min} / \|\|C\|\|_\infty),$$

*or $\infty$ if there are no constraints[2.12]. Then for any $x \in C(0)$,*

1. *For any $\delta > 0, y \in \mathbb{R}_+^m$, if $\|y - x\|_\infty \leqslant \delta \rho_\infty$, then $y \in C(\delta)$.*
2. *In particular, if $\delta \geqslant 1/(2\rho_\infty)$, the integral vector $[x]$ is in $C(\delta)$.*

Now let $s_1(\delta), s_2(\delta)$ generalize the $s_1, s_2$ of (2.10). By considering the $x \in C(0)$ that achieve $s_1, s_2$ it follows from part 1 of the proposition that

$$s_1(\delta) \leqslant s_1 - m \delta \rho_\infty, \qquad s_2(\delta) \geqslant s_2 + m \delta \rho_\infty, \tag{2.17}$$

which we will use in §4.

# 3 Relationship to I-divergence

The relative entropy or divergence $D(\cdot \| \cdot)$ between density vectors has a well-known generalization to arbitrary non-negative vectors

$$\mathcal{D}(u \| v) \triangleq \sum_i u_i \ln \frac{u_i}{v_i} - \sum_i u_i + \sum_i v_i, \qquad u, v \in \mathbb{R}_+^m, \tag{3.1}$$

---

2.11. This definition of $\nu^*$ is much simpler than the one in [Oik17].

2.12. Recall that the infinity norm $\|\| \cdot \|\|_\infty$ of a matrix is the maximum of the $\ell_1$ norms of the rows.

known as *I-divergence* ([Csi91], [Csi96]), and sometimes as *generalized KL-divergence*; it reduces to $D(u\|v)$ when $u,v$ are density vectors. Minimizing this function is extensively used for inference or approximation ([KF09], [Mur12]). It also appears in the *entropic prior* for non-negative real vectors [Ski89], [SS06], [vDv14]. The minimization of $\mathcal{D}(u\|v)$ with respect to $u$ is known as 'I-projection', and with respect to $v$ as 'M-projection' ('information' and 'moment' projection respectively). These two minimizations produce qualitatively different results ([KF09], Ch. 8, [Mur12], Ch. 21), and M-projection does not have the axiomatic justification [Ski89], [Csi91], [Csi96] that I-projection does. Nevertheless, it is quite commonly used in applications involving density vectors. In this section we establish a relationship between minimizing $\mathcal{D}(u\|v)$ w.r.t. $u$, to which we refer as MINIDIV, and maximizing $G(x\|y)$ w.r.t. $x$, which we call MAXGRENT.

First we observe that given a prior $\mu$, if $u$ and $v$ have the same sum, the difference between the values of $G$ at $u,v$ is the same as that of $\mathcal{D}$:
$$G(u\|\mu) - G(v\|\mu) = \mathcal{D}(v\|\mu) - \mathcal{D}(u\|\mu), \quad \text{if} \quad \sum_i u_i = \sum_i v_i.$$

Second, since $\mathcal{D}(u\|v)$ is a convex function of $u$ and it is bounded from below by 0, if there is a $u^*$ and multipliers $\rho^*, \sigma^*$ such that
$$\begin{aligned} Au^* &= b, \\ u_j^* &= v_j e^{-(\rho^* \cdot A_{\cdot j} + \sigma^* \cdot C_{\cdot j})}, \quad 1 \leq j \leq m, \\ \sigma_k^*(C_{\cdot k} \cdot u^* - d_k) &= 0, \quad \sigma_k^* \geq 0, \quad 1 \leq k \leq l_I, \end{aligned} \quad (3.2)$$

then $u^*$ minimizes $\mathcal{D}(u\|v)$ under the constraints $Au = b, Cu \leq d$. The similarity of the expression for $u_j^*$ in (3.2) to that for $x_j^*$ in (2.8) suggests some relationship between the two. Indeed, by using suitable priors, and in one case adding a constraint, the MAXGRENT and MINIDIV (I-projection) methods can be made to produce the same solution:

**Proposition 3.1.** *Consider linear constraints on a vector in $\mathbb{R}_+^m$ of the type in (2.7), and let $x^* = x^*(y)$ be the solution of the MAXGRENT problem with prior $y$ and these constraints on $x$. Likewise, let $u^* = u^*(v)$ be the solution to the MINIDIV problem with prior $v$ and the same constraints on $u$. Then*

1. *Using the prior $\tilde{v} = (\sum_i x_i^*) y$ in MINIDIV makes $u^*(\tilde{v}) = x^*(y)$.*
2. *Using the prior $\tilde{y} = v / \sum_i u_i^*$ in MAXGRENT, and adding the constraint $\sum_i x_i = \sum_i u_i^*$, makes $x^*(\tilde{y}) = u^*(v)$.*

The proposition shows a certain correspondence, but must be interpreted with care: it does not imply that the two methods are 'equivalent'. Part 1 does *not* say that "for any problem that MAXGRENT can solve MINIDIV can get the same solution, if only one uses the right prior"; finding the "right prior" for MINIDIV requires solving the problem by MAXGRENT (we only need the sum of the MAXGRENT solution). An analogous comment applies to part 2: the prior to be used in MAXGRENT requires having the MINIDIV solution, or at least its sum.

There is also a fundamental conceptual difference: in our discrete setting, the classical MAXENT method can be viewed as always maximizing the uncertainty, equivalently the number of realizations, and classical MINXENT as minimizing a directed distance from the prior; when the prior expresses maximal uncertainty (is uniform) this coincides with maximizing the uncertainty. Our MAXGRENT *always* maximizes the number of realizations (uncertainty) *irrespective* of the prior[3.1] whereas MINIDIV behaves like MINXENT but also applies to non-density vectors.

---

3.1. In §7 where the prior is a density vector, the number of realizations becomes the *probability*.

An equally important difference has to do with *concentration*, which is the subject of the next few sections: I-divergence does *not* concentrate in our combinatorial sense, i.e. by maximizing the number of realizations of the count vector that minimizes it. [Csi96] §7 points out that such an interpretation *does not seem to apply* to I-divergence. If I-divergence doesn't concentrate, neither do its many *generalizations*, e.g. the $\alpha, \beta$-divergences of [CCA11]. However its specialization $D(\cdot\|\cdot)$ to density vectors does concentrate, [Gr8].

# 4 Concentration around the maximum relative entropy value

In this section we show that with suitable parameters $\delta, \varepsilon, \eta$, the MAXGRENT vector $\nu^*$ dominates with respect to the number of realizations $\#_\mu$ the entire set of count vectors that have entropy $\eta$-far from $G^*$. This concentration result is elaborated further in §6.

For a given $n$, we partition the domain $C(\delta)$ of (2.16) into two sets, defined by proximity to the maximum *value* of the relative entropy:

$$\begin{aligned} \mathcal{A}_n(\delta, \eta) &\triangleq \{\nu \in N_n \cap C(\delta) : G(\nu \| \mu) \geqslant (1-\eta) G^*(\mu)\} \\ \mathcal{B}_n(\delta, \eta) &\triangleq \{\nu \in N_n \cap C(\delta) : G(\nu \| \mu) < (1-\eta) G^*(\mu)\} \end{aligned}, \quad \eta \in (0,1). \quad (4.1)$$

Next, we want an approximation (subset) of the set of all integral vectors in $C(\delta)$; for simplicity, it should be independent of $\delta$. We do this by bounding the sums of these vectors by

$$n_1 \triangleq \lceil s_1 \rceil - m/2, \quad n_2 \triangleq \lceil s_2 \rceil + m/2, \quad (4.2)$$

which we get by putting $\delta = 1/2\rho_\infty$ in (2.17); then Proposition 2.10(2) says that $\nu \in \mathbb{N}^m$ with sums in $[n_1, n_2]$ are in $C(\delta)$, and ensures that $n_1 \geqslant 0$. We are interested in the (disjoint) unions of the sets (4.1) over $n \in \{n_1, \ldots, n_2\}$:

$$\begin{aligned} \mathcal{A}_{n_1:n_2}(\delta, \eta) &\triangleq \{\nu : \sum_i \nu_i = n, n_1 \leqslant n \leqslant n_2, \nu \in C(\delta), G(\nu \| \mu) \geqslant (1-\eta) G^*(\mu)\}, \\ \mathcal{B}_{n_1:n_2}(\delta, \eta) &\triangleq \{\nu : \sum_i \nu_i = n, n_1 \leqslant n \leqslant n_2, \nu \in C(\delta), G(\nu \| \mu) < (1-\eta) G^*(\mu)\}. \end{aligned} \quad (4.3)$$

The union of these two sets equals $N_{n_1:n_2} \cap C(\delta)$. We will derive an upper bound on the number of realizations of $\mathcal{B}_{n_1:n_2}(\delta, \eta)$ and then a lower bound on the number of realizations of $\mathcal{A}_{n_1:n_2}(\delta, \eta)$. [The number of realizations of a set $X$ of count vectors is simply $\#X = \sum_{\nu \in X} \#\nu$.]

Eq. (4.7) of [Oik17] bounds the multinomial coefficient in terms of $G(\nu)$:

$$e^{-\frac{m}{12}} S(\nu) e^{G(\nu)} \leqslant \binom{\nu_1 + \cdots + \nu_m}{\nu_1, \ldots, \nu_m} \leqslant S(\nu) e^{G(\nu)}, \quad \text{where} \quad S(\nu) \triangleq \frac{\sqrt{n}}{(2\pi)^{(m-1)/2}} \frac{1}{\sqrt{\nu_1 \cdots \nu_m}}, \quad (4.4)$$

where we assume that all $\nu_i > 0$, recall Remark 2.4. Multiplying all sides of the inequality by $\mu_1^{\nu_1} \cdots \mu_m^{\nu_m}$ and using (2.1) and the middle equation in (2.4) we bound $\#_\mu \nu$:

$$e^{-\frac{m}{12}} S(\nu) e^{G(\nu \| \mu)} \leqslant \#_\mu \nu \leqslant S(\nu) e^{G(\nu \| \mu)}. \quad (4.5)$$

Therefore

$$\#_\mu \mathcal{B}_n(\delta, \eta) \leqslant \sum_{\nu \in \mathcal{B}_n(\delta, \eta)} S(\nu) e^{G(\nu \| \mu)} < e^{(1-\eta) G^*(\mu)} \sum_{\nu \in N_n} S(\nu),$$

and so
$$\#_\mu \mathcal{B}_{n_1:n_2}(\delta,\eta) = \sum_{n_1 \leqslant n \leqslant n_2} \#_\mu \mathcal{B}_n(\delta,\eta) < e^{(1-\eta)G^*(\mu)} \sum_{n_1 \leqslant n \leqslant n_2} \sum_{\nu \in N_n} S(\nu).$$

The bound
$$\sum_{n_1 \leqslant n \leqslant n_2} \sum_{\nu \in N_n} S(\nu) \leqslant \frac{\sqrt{\pi}}{(m+1)\, 2^{(m-3)/2}\, \Gamma(m/2)} \\ \left((\sqrt{s_2 + m/2 + 2} + \sqrt{m})^{m+1} - (\sqrt{s_1 - m/2} + \sqrt{m})^{m+1}\right) \\ \triangleq C_1(s_1, s_2) \tag{4.6}$$

was derived in §4.2 of [Oik17]. Using it in the expression for $\#_\mu \mathcal{B}_{n_1:n_2}(\delta,\eta)$ above,
$$\#_\mu \mathcal{B}_{n_1:n_2}(\delta,\eta) < C_1(s_1, s_2)\, e^{(1-\eta)G^*(\mu)}. \tag{4.7}$$

Now for the lower bound on $\#_\mu \mathcal{A}_{n_1:n_2}(\delta,\eta)$ we use the lower bound (4.5) on $\#_\mu \nu^*$:
$$\#_\mu \nu^* \geqslant S(\nu^*)\, e^{-m/12}\, e^{G(\nu^* \| \mu)}. \tag{4.8}$$

We also have bounds on $G(y \| \mu)$ when $y$ is inside a hypercube centered at $x$ and of side $2\zeta$:

**Proposition 4.1.** *Given $\zeta > 0$ and $x, y \in \mathbb{R}_+^m$, if $\|y - x\|_\infty \leqslant \zeta$,*
$$-g(x)\zeta - \frac{1}{2} h(x,\zeta)\, \zeta^2 \leqslant G(y\|\mu) - G(x\|\mu) \leqslant g(x)\zeta - \frac{1}{2} h(x,-\zeta)\, \zeta^2$$

*where $g(x) \triangleq \sum_i \ln \frac{\mu_i}{\chi_i}$, $h(x,\zeta) \triangleq \sum_i \frac{1}{x_i - \zeta} - \frac{m}{\|x\|_1/m - \zeta}$, and assuming that $x > (\zeta, \ldots, \zeta)$ for the lower bound. Further, $h(x,\zeta) > 0$ unless all $x_i$ are equal, in which case it is 0.*

Since $\|\nu^* - x^*\|_\infty \leqslant 1/2$ by Proposition 2.9, taking $x = x^*, y = \nu^*, \zeta = 1/2$ in the lower bound of Proposition 4.1 we obtain
$$x^* > 1/2 \implies G(\nu^* \| \mu) \geqslant G(x^* \| \mu) - \frac{1}{2} \sum_i \ln \frac{\mu_i}{\chi_i^*} - \frac{1}{8}\left(\sum_i \frac{1}{x_i^* - 1/2} - \frac{m}{s^*/m - 1/2}\right). \tag{4.9}$$

Putting (4.9) in (4.8) together with a lower bound on $S(\nu^*)$ obtained by using $\nu_i^* \leqslant x_i^* + 1/2$ in (4.4), we find that if $x^* > 1/2$
$$\#_\mu \nu^* \geqslant e^{G^*(\mu)} \frac{e^{-m/12} \sqrt{s^*}}{(2\pi)^{(m-1)/2}} \frac{1}{\prod_i \sqrt{\mu_i}} \prod_{i=1}^m \sqrt{\frac{\chi_i^*}{x_i^* + 1/2}}\, e^{-\frac{1}{8}\left(\sum_{i=1}^m \frac{1}{x_i^* - 1/2} - \frac{m}{s^*/m - 1/2}\right)} \\ \triangleq C_0(x^*, \mu)\, e^{G^*(\mu)} \tag{4.10}$$

where $G^*(\mu)$ is defined in (2.9). Finally from Proposition 2.10 and (4.9), and from (4.10) and (4.7), we obtain a lower bound on the ratio of the number of realizations of $\nu^*$ to that of the set $\mathcal{B}_{n_1:n_2}(\delta,\eta)$:

**Lemma 4.2.** *Let $x^*$ be the solution to the maximization problem (2.7) with $y = \mu \geqslant 1$, assume that $x^* > 1/2$, and let $G^*(\mu)$ be the maximum generalized relative entropy (2.9). Then for any $\delta > 0$ and $\eta \in (0,1)$*
$$\frac{\#_\mu \nu^*}{\#_\mu \mathcal{B}_{n_1:n_2}(\delta,\eta)} > \frac{(m+1)\,\Gamma(m/2)\, e^{-m/12}}{2\pi^{m/2}} \frac{1}{\prod_i \sqrt{\mu_i}} \frac{C_2(x^*)\, C_4(x^*)}{C_3(s_1, s_2)}\, e^{\eta G^*(\mu)},$$

where $\mathcal{B}_{n_1:n_2}(\delta,\eta)$ is defined in (4.3), and the constants are

$$C_2(x^*) \triangleq \left(s^* \prod_{i=1}^m \frac{x_i^*}{x_i^* + 1/2}\right)^{1/2},$$

$$C_3(s_1,s_2) \triangleq \left(\sqrt{s_2 + m/2 + 2} + \sqrt{m}\right)^{m+1} - \left(\sqrt{s_1 - m/2} + \sqrt{m}\right)^{m+1},$$

$$C_4(x^*) \triangleq e^{-\frac{1}{8}\left(\sum_{i=1}^m \frac{1}{x_i^* - 1/2} - \frac{m}{s^*/m - 1/2}\right)},$$

where $s^* \triangleq \sum_i x_i^*$. Lastly, if $\delta,\eta$ satisfy

$$\delta \geq \frac{1}{2\rho_\infty}, \quad \eta \geq \frac{1}{2G^*(\mu)} \sum_{1 \leq i \leq m} \ln \frac{\mu_i}{\chi_i^*} + \frac{1}{8G^*(\mu)} \left(\sum_{1 \leq i \leq m} \frac{1}{x_i^* - 1/2} - \frac{m}{s^*/m - 1/2}\right),$$

where $\rho_\infty$ is defined in Proposition 2.10, the count vector $\nu^* = [x^*]$ is in the set $\mathcal{A}_{n^*}(\delta,\eta)$ of (4.1).

The $C_2, C_4$ above and the $C_0$ in (4.10) are related by $C_0 = e^{-m/12}(2\pi)^{-(m-1)/2} \prod_i \mu_i^{-1/2} C_2 C_4$. The second claim of the theorem follows from (4.1) and the last line of (4.9).

The dependence of the lower bound on the constraints of the problem is implicit, via $G^*$ and $\rho_\infty$. The constant $C_3(s_1,s_2)$ shows that, as intuitively expected, the bound becomes weaker as the range of the allowable sum of $x$ increases.

## 5 Concentration around the maximum relative entropy vector

In this section we establish a lower bound on the ratio of the number of realizations of $\nu^*$ to that of the set $\mathcal{B}_{n_1:n_2}$ analogous to that of §4, but exponential in (the square of) a parameter $\vartheta$ that governs the $\ell_\infty$ distance of a vector from $x^*$.

For fixed $n$, we partition the set $N_n \cap C(\delta)$ of count vectors that satisfy the constraints into two subsets, according to the distance from the optimal vector $x^*$ measured by the $\ell_\infty$ (max) norm:

$$\begin{aligned} \mathcal{A}_n(\delta,\vartheta) &\triangleq \{\nu \in N_n \cap C(\delta) : \|\nu - x^*\|_\infty \leq \vartheta \|x^*\|_\infty\} \\ \mathcal{B}_n(\delta,\vartheta) &\triangleq \{\nu \in N_n \cap C(\delta) : \|\nu - x^*\|_\infty > \vartheta \|x^*\|_\infty\} \end{aligned}, \quad \vartheta > 0. \quad (5.1)$$

These definitions are both significantly stronger, and more intuitively-satisfying than the ones given in [Oik17]. Because both $\nu$ and $x^*$ are $\geq 0$, it is always the case that $\|\nu - x^*\|_\infty \leq \max(\|\nu\|_\infty, \|x^*\|_\infty)$.

The (disjoint) unions of the sets (5.1) over $n \in [n_1, n_2]$, defined in (4.2), are

$$\begin{aligned} \mathcal{A}_{n_1:n_2}(\delta,\vartheta) &\triangleq \{\nu : \sum_i \nu_i = n, n_1 \leq n \leq n_2, \nu \in C(\delta), \|\nu - x^*\|_\infty \leq \vartheta \|x^*\|_\infty\}, \\ \mathcal{B}_{n_1:n_2}(\delta,\vartheta) &\triangleq \{\nu : \sum_i \nu_i = n, n_1 \leq n \leq n_2, \nu \in C(\delta), \|\nu - x^*\|_\infty > \vartheta \|x^*\|_\infty\}. \end{aligned} \quad (5.2)$$

To understand the geometry of $G(x\|\mu)$ around $x^*$ consider the sets

$$K^*(\vartheta) \triangleq \{x : \|x - x^*\|_\infty \leq \vartheta \|x^*\|_\infty\}, \quad D^*(\vartheta) \triangleq \{x : |\sum_i x_i - s^*| \leq \vartheta \|x^*\|_\infty\}, \quad (5.3)$$

a hypercube centered at $x^*$, and the region in the first orthant between the two hyperplanes $\sum_i x_i = s^* \pm \vartheta \|x^*\|_\infty$. Their importance is that

**Proposition 5.1.** *Let $x^*$ be the unique maximizer of $G(x\|\mu)$ over a compact convex set $C$. Then*
 1. *For every $x \in C$ outside $K^*(\vartheta)$, there is a $y \in C$ and on the surface of $K^*(\vartheta)$ s.t. $G(y\|\mu) > G(x\|\mu)$.*

2. *For every $x \in C$ outside $D^*(\vartheta)$, there is a $y \in C$ and in $D^*(\vartheta)$ s.t. $G(y \| \mu) > G(x \| \mu)$.*

The proposition applies to the polyhedral set $C(\delta)$, $\delta \geqslant 0$ [5.1], and its additional generality will be useful later. Part 1 tells us that to maximize $G(x \| \mu)$ over $B_n(\delta, \vartheta)$ we can restrict our attention to points in $C(\delta)$ that lie on the *surface* of $K^*(\vartheta)$. Whereas neither $B_n(\delta, \vartheta)$ nor this set are convex, the latter is easier to deal with. So the proposition provides the starting point for a central result of this section, again valid for more general sets than $C(\delta)$:

**Lemma 5.2.** *Let $C$ be a compact convex set, let $x^*$ be the unique maximizer[5.2] of $G(x \| \mu)$ over $C$, assume that $\|x^*\|_\infty < 2 \|x^*\|_1$, and fix $\vartheta \in \left(0, \frac{1}{2} \frac{\|x^*\|_1}{\|x^*\|_\infty} - 1\right]$. Then for any $x \in C$ with $\|x - x^*\|_\infty > \vartheta \|x^*\|_\infty$,*

$$G(x^* \| \mu) - G(x \| \mu) \geqslant \frac{\vartheta^2}{2(1+\vartheta)} \frac{1}{1 - (1+\vartheta) \|x^*\|_\infty / \|x^*\|_1} \|x^*\|_\infty.$$

The applicability of the lemma is restricted by the condition $\|x^*\|_\infty < 2 \|x^*\|_1$; clearly one can formulate problems where $\|x^*\|_1 / \|x^*\|_\infty \leqslant 2$, e.g. the one in Example 2.1. Otherwise, it can be shown that for a fixed ratio $\|x^*\|_\infty / \|x^*\|_1$ the coefficient of $\|x^*\|_\infty$ in the lower bound of the lemma increases monotonically with $\vartheta \in \left(0, \frac{1}{2} \frac{\|x^*\|_1}{\|x^*\|_\infty} - 1\right]$. Lemma 2.6 provides conditions for the uniqueness of $x^*$ and then Lemma 5.2 applies to $C(\delta)$, $\delta \geqslant 0$. Its proof is in Appendix B, and the restriction on $x^*$ is (B.14) there; it might be possible to weaken this restriction with more work on the proof. Despite the fact that $\mu$ does not appear in the bound, it is implicit in $x^*$.

Lemma 5.2 says that if $x$ is far enough from $x^*$ then $G(x \| \mu)$ is far from $G(x^* \| \mu)$, the distances being measured by $\|x^*\|_\infty$. To provide some context around this, if $x^*$ maximizes a function $G$ over a set $C$, a statement that there is a $\kappa > 0$ s.t. $G(x^*) - G(x) \geqslant \kappa \|x - x^*\|$ is known as a *growth condition* for $G$ around $x^*$, and the related statement that there is a $\kappa'$ s.t. $\|x - x^*\| \leqslant \kappa' (G(x^*) - G(x))$ is known as an *error bound* on $x$. In both statements $x$ may be restricted to lie in a subset of $C$; see [FP03] for a detailed treatment. With this in mind, the following negative results help in assessing that of the lemma:

**Proposition 5.3.** *Let $C \subset \mathbb{R}^m_+$ be a convex set, $x^*$ be the maximizer of $G(x \| \mu)$ over $C$, and let $\|\cdot\|$ be the $\ell_\infty$, $\ell_1$, or $\ell_2$ norm. Then*

1. *Given $\xi > 0$, there is no constant $\kappa > 0$ such that the growth condition*

$$G(x^* \| \mu) - G(x \| \mu) \geqslant \kappa \|x - x^*\|$$

   *holds for all $x \in C$ with $\|x - x^*\| \leqslant \xi$.*

2. *Given $\vartheta > 0$, there is no $\kappa > 0$ independent of $\vartheta$ such that*

$$G(x^* \| \mu) - G(x \| \mu) \geqslant \kappa \vartheta \|x^*\|$$

   *holds for all $x \in C$ with $\|x - x^*\| \geqslant \vartheta \|x^*\|$.*

Returning to Lemma 5.2, to proceed as as in §4 and derive a lower bound on the ratio $\#_\mu v^* / \#_\mu \mathcal{B}_{n_1:n_2}(\delta, \vartheta)$ we need the maximum value of $G(x \| \mu)$ over $C(\delta)$, which may be larger than the one over $C(0)$. This was not a problem in §4, because such larger values belonged by definition to the 'good' set $\mathcal{A}_n(\delta, \eta)$ of (4.1), but here we cannot guarantee that the maximizer of $G$ over $C(\delta)$ is in some $\mathcal{A}_n(\delta, \vartheta)$, as defined in (5.1) in terms of $x^*$. An upper bound on $\max_{x \in C(\delta)} G(x \| \mu)$ suffices for our purposes.

---

[5.1]. and actually holds for any $\ell_p$ norm, see the proof.

[5.2]. Here we are re-using the $x^*$ of (2.9) with an extended meaning, to avoid introducing more new notation.

To derive the bound we view $C(\delta)$ as a *additive perturbation* of $C(0)$ parameterized by $\tilde{\beta}$

$$C(\delta,\tilde{\beta}) = \{x \in \mathbb{R}^m_+ | Ax = b + \delta\tilde{\beta}, Cx \leq d + \delta\tilde{d}\}, \qquad -\tilde{b} \leq \tilde{\beta} \leq \tilde{b},$$

so $x \in C(\delta)$ iff $\exists \tilde{\beta}: x \in C(\delta,\tilde{\beta})$. A well-known result (e.g. [BV04] §5.2.6) relates the solution of the perturbed problem $\max_{x \in C(\delta,\tilde{\beta})} G(x\|\mu)$ to that of the original, unperturbed problem $\max_{x \in C(0)} G(x\|\mu)$ via the *dual* of the unperturbed problem: if strong duality holds and the solution of the dual is $(\lambda^*, \zeta^*, G^*)$, then the optimal value of the perturbed primal obeys

$$\max_{x \in C(\delta,\tilde{\beta})} G(x\|\mu) \leq G^*(\mu) + \delta\lambda^* \cdot \tilde{\beta} + \delta\zeta^* \cdot \tilde{d}$$

for all perturbations $\delta\tilde{\beta}, \delta\tilde{d}$ for which the perturbed primal is feasible. In our case $\lambda^*, \zeta^*$ were found in §2.4 and the perturbed primal has the same form as the unperturbed one, so we know from §2.4 that strong duality holds. It follows from the above that with

$$\tilde{G} \triangleq |\lambda^*| \cdot \tilde{b} + \zeta^* \cdot \tilde{d} \tag{5.4}$$

we have

$$\max_{x \in C(\delta)} G(x\|\mu) \leq G^*(\mu) + \delta\tilde{G}, \qquad \text{where} \quad \tilde{G} \geq G^*(\mu) \tag{5.5}$$

Now we can proceed as in §4 after (4.5): to bound $\#_\mu \mathcal{B}_n(\delta, \vartheta)$ we apply Lemma 5.2 to $C(0)$, and using (5.5) we get

$$\#_\mu \mathcal{B}_n(\delta, \vartheta) = \sum_{\nu \in \mathcal{B}_n(\delta,\vartheta)} \#_\mu(\nu) \leq e^{G^*(\mu) + \delta\tilde{G} - \vartheta^2 K(\vartheta, x^*)\|x^*\|_\infty} \sum_{\nu \in N_n} S(\nu)$$

where

$$K(\vartheta, x^*) \triangleq \frac{1}{2(1+\vartheta)\left(1 - (1+\vartheta)\|x^*\|_\infty / \|x^*\|_1\right)}, \tag{5.6}$$

and again continuing as in §4 we have an analogue of (4.7):

$$\#_\mu \mathcal{B}_{n_1:n_2}(\delta, \vartheta) \leq C_1(s_1, s_2) e^{G^*(\mu) + \delta\tilde{G} - \vartheta^2 K(\vartheta, x^*)\|x^*\|_\infty}. \tag{5.7}$$

Combining (5.7) with (4.10) we get a lower bound on the ratio $\#_\mu \nu^* / \#_\mu \mathcal{B}_{n_1:n_2}(\delta, \vartheta)$ analogous to that of Lemma 4.2. However here the parameters $\delta$ and $\vartheta$ are intertwined, unlike $\delta$ and $\eta$ which were independent:

**Lemma 5.4.** *Let $x^*$ be the solution to the maximization problem* (2.7) *with $y = \mu \geq 1$. Assuming that $x^* > 1/2$ and $\|x^*\|_\infty < 2\|x^*\|_1$, if $\delta, \vartheta$ are such that*

$$\delta \geq \frac{1}{2\rho_\infty}, \quad \vartheta \geq \frac{1}{2\|x^*\|_\infty},$$

*where $\rho_\infty$ is defined in Proposition* 2.10, *the vector $\nu^* = [x^*]$ is in the set $\mathcal{A}_{n^*}(\delta, \vartheta)$ of* (5.1). *Further, if*

$$\vartheta \leq \frac{1}{2}\frac{\|x^*\|_1}{\|x^*\|_\infty} - 1, \qquad \delta < \vartheta^2 K(\vartheta, x^*) \frac{\|x^*\|_\infty}{\tilde{G}},$$

*where $\tilde{G}$ and $K(\vartheta, x^*)$ are defined in* (5.4) *and* (5.6), *then*

$$\frac{\#_\mu \nu^*}{\#_\mu \mathcal{B}_{n_1:n_2}(\delta, \vartheta)} \geq \frac{(m+1)\Gamma(m/2)e^{-m/12}}{2\pi^{m/2}} \frac{1}{\prod_i \sqrt{\mu_i}} \frac{C_2(x^*)C_4(x^*)}{C_3(s_1, s_2)} e^{\vartheta^2 K(\vartheta, x^*)\|x^*\|_\infty - \delta\tilde{G}},$$

*where the set $\mathcal{B}_{n_1:n_2}(\delta, \vartheta)$ is defined in* (5.2), *and the constants $C_2, C_3, C_4$ are as in Lemma* 4.2.

Apart from the exponential factor the lower bound is identical to that of Lemma 4.2, but is subject to the restriction $\|x^*\|_\infty < 2\|x^*\|_1$. It shows that the number of realizations of the vector $v^*$, which does not belong to $\mathcal{C}(0)$ but to $\mathcal{C}(\delta)$ with $\delta > 0$, dominates that of the set $\mathcal{B}_{n_1:n_2}(\delta, \vartheta)$. The first set of conditions on $\delta$ and $\vartheta$ ensures that $v^* \in \mathcal{A}_{n^*}(\delta, \vartheta)$ by Propositions 2.9 and 2.10. The second set comes from Lemma 5.2, and of these the condition on $\delta$ is not needed for the bound to hold, but for it to be useful: if $\delta$ is too large, the maximizer of $G(x \| \mu)$ over $\mathcal{C}(\delta)$ may be far from $x^*$ so we cannot expect concentration to occur in a region of size $\vartheta \|x^*\|_\infty$ around $x^*$.

# 6 Large problems: scaling and concentration

To study concentration in 'large' problems we regard the matrices $A, C$ as defining the *structure* of the problem's constraints and the vectors $b, d$ as defining their *data* or *values*. So here we assume that we have a problem with a given, fixed structure, but whose *size*, defined as that of the elements of the data, can become as large as we please while the structure remains invariant. Example 2.1 shows a very simple case, and Example 8.1 a more complex one. Recall that Lemmas 4.2 and 5.4 apply to problems of a given structure *and* given size.

We will consider the simplest way of increasing the problem's size: multiplying $b, d$ by a factor $c \geqslant 1$, a process we call *scaling*; its properties are described in Proposition 2.7. In this section we derive a *concentration threshold* $\hat{c} = \hat{c}(\delta, \varepsilon, \eta, \vartheta)$, which is such that if the problem is scaled by a factor $c \geqslant \hat{c}$, concentration around $G^*$ or $v^*$ will occur to the degree specified by $\varepsilon$: the number of realizations of $v^*$ will dominate by the factor $1/\varepsilon$ the realizations of the set of all vectors whose generalized relative entropy is $\eta$-far from $G^*$ or that are $\vartheta$-far from $x^*$ w.r.t. $\ell_\infty$ distance. This is Theorem 6.1 below.

## 6.1 The effects of scaling

We look at how the results of Lemmas 4.2 and 5.4 change when the original problem is scaled by $c \geqslant 1$. We begin with the constants and exponents.

The constants $C_2, C_3, C_4$ in Lemma 4.2 are common to both lemmas and behave as follows. With $x^* \stackrel{c}{\mapsto} cx^*$ from Proposition 2.7,

$$C_2(cx^*) = \frac{\sqrt{s^*}}{c^{m/2-1}} \left( \prod_{1 \leqslant i \leqslant m} \frac{\chi_i^*}{x_i^* + 1/(2c)} \right)^{1/2} \geqslant \frac{\sqrt{s^*}}{c^{m/2-1}} \left( \prod_{1 \leqslant i \leqslant m} \frac{\chi_i^*}{x_i^* + 1/2} \right)^{1/2}, \quad c \geqslant 1, \tag{6.1}$$

since $\chi^*$ is invariant under scaling, and the first product above increases as $c \nearrow$. Next, for $C_4(cx^*)$,

$$-\frac{1}{8} \left( \sum_{1 \leqslant i \leqslant m} \frac{1}{cx_i^* - 1/2} - \frac{m}{cs^*/m - 1/2} \right) > -\frac{1}{8c} \left( \sum_{1 \leqslant i \leqslant m} \frac{1}{x_i^* - 1/2} - \frac{m^2}{s^*} \right)$$

since $x^* > 1/2$ and $c \geqslant 1$. This implies that

$$C_4(cx^*) > e^{-\frac{1}{8c} \left( \sum_{i=1}^m \frac{1}{x_i^* - 1/2} - \frac{m^2}{s^*} \right)}. \tag{6.2}$$

For $C_3$ we have

$$C_3(cs_1, cs_2) = c^{(m+1)/2} \left( \left( \sqrt{s_2 + m/2c + 2/c} + \sqrt{m/c} \right)^{m+1} - \left( \sqrt{s_1 - m/2c} + \sqrt{m/c} \right)^{m+1} \right)$$

and the function of $c$ multiplying $c^{(m+1)/2}$ above decreases as $c \nearrow$ [6.1], so

$$C_3(cs_1, cs_2) \leqslant c^{(m+1)/2}\left(\left(\sqrt{s_2 + m/2 + 2} + \sqrt{m}\right)^{m+1} - \left(\sqrt{s_1 - m/2} + \sqrt{m}\right)^{m+1}\right), \quad c \geqslant 1. \tag{6.3}$$

For the exponents in the two lemmas we have, respectively,

$$\begin{aligned}\eta G^*(\mu) &\overset{c}{\mapsto} c\eta G^*(\mu), \\ \vartheta^2 K(\vartheta, x^*)\|x^*\|_\infty - \delta\tilde{G} &\overset{c}{\mapsto} c\left(\vartheta^2 K(\vartheta, x^*)\|x^*\|_\infty - \delta\tilde{G}\right).\end{aligned} \tag{6.4}$$

The first follows from Proposition 2.7. The second follows from $\tilde{G} \overset{c}{\mapsto} c\tilde{G}$, as (2.13) shows that the dual solution $\lambda^*, \xi^*$ is invariant under scaling, and from the fact that $K(\vartheta, x^*)$ is likewise invariant.

Now we look at the conditions on $\delta, \eta$ in Lemma 4.2. These will hold after scaling if

$$\delta \geqslant \frac{1}{2c\rho_\infty}, \quad \eta \geqslant \frac{1}{2cG^*(\mu)} \sum_{1 \leqslant i \leqslant m} \ln \frac{\mu_i}{\chi_i^*} + \frac{1}{8c^2 G^*(\mu)} \left(\sum_{1 \leqslant i \leqslant m} \frac{1}{x_i^* - 1/2} - \frac{m^2}{s^*}\right).$$

[For the first condition, Proposition 2.10 shows that $\tilde{b} \overset{c}{\mapsto} c\tilde{b}$ implies $\rho_\infty \overset{c}{\mapsto} c\rho_\infty$, and for the second condition we used the simplification that led to (6.2).] These bounds are equivalent to the scaling factor satisfying

$$c \geqslant \frac{1}{2\delta\rho_\infty}, \quad c \geqslant \frac{1}{2\eta G^*(\mu)} \sum_{1 \leqslant i \leqslant m} \ln \frac{\mu_i}{\chi_i^*} + \frac{1}{8c\eta G^*(\mu)} \sum_{1 \leqslant i \leqslant m} \left(\frac{1}{x_i^* - 1/2} - \frac{m^2}{4s^*}\right) \tag{6.5}$$

where the 2nd condition leads to a quadratic in $c$.

For Lemma 5.4, the conditions for $\delta$ and $\vartheta$ will hold after scaling if

$$\vartheta \leqslant \frac{1}{2}\frac{\|x^*\|_1}{\|x^*\|_\infty} - 1, \quad \delta < \vartheta^2 K(\vartheta, x^*)\frac{\|x^*\|_\infty}{\tilde{G}}, \tag{6.6}$$

whose right-hand sides are invariant under scaling, and if

$$c \geqslant \frac{1}{2\vartheta\|x^*\|_\infty}, \quad c \geqslant \frac{1}{2\delta\rho_\infty}. \tag{6.7}$$

## 6.2 The concentration threshold

Putting (6.1) to (6.7) together we see that whether we consider concentration in the sense of §4 around $G^*$ or in the sense of §5 around $x^*$, when the problem data $b$ is scaled by $c \geqslant 1$, there is a lower bound on the ratio $\#_\mu v^* / \#_\mu \mathcal{B}_{n_1:n_2}$ of the form $Kc^{-\beta}e^{\alpha c}$, where the constants are

$$K \triangleq \frac{(m+1)\Gamma(m/2) e^{-m/12}}{2\pi^{m/2}\sqrt{\mu_1 \cdots \mu_m}} \frac{\left(s^* \prod_{1 \leqslant i \leqslant m} \frac{\chi_i^*}{x_i^* + 1/2}\right)^{1/2} \exp\left(-\frac{1}{8}\left(\sum_{i=1}^m \frac{1}{x_i^* - 1/2} - \frac{m^2}{s^*}\right)\right)}{\left(\sqrt{s_2 + m/2 + 2} + \sqrt{m}\right)^{m+1} - \left(\sqrt{s_1 - m/2} + \sqrt{m}\right)^{m+1}},$$

$$\alpha \triangleq \begin{cases} \eta G^*, & G^*\text{-case} \\ \vartheta^2 K(\vartheta, x^*)\|x^*\|_\infty - \delta\tilde{G}, & x^*\text{-case} \end{cases}, \quad \beta \triangleq m - 1/2, \quad \gamma \triangleq \ln\frac{1}{\varepsilon K}. \tag{6.8}$$

Provided that $\alpha > 0$, $Kc^{-\beta}e^{\alpha c}$ will increase without limit as $c$ increases, establishing the concentration phenomenon. To ensure that $Kc^{-\beta}e^{\alpha c} \geqslant 1/\varepsilon$ for some $\varepsilon > 0$, $c$ must satisfy $\alpha c - \beta \ln c - \gamma = 0$.

---

[6.1]. After some algebra, its derivative can be shown to be negative if $s_2 \geqslant s_1$.

It is shown in the Appendix that if $f(c) = \alpha c - \beta \ln c - \gamma$ with $\alpha, \beta, \gamma$ as in (6.8), then $\alpha, \gamma > 0$, and $Kc^{-\beta}e^{\alpha c} \geq 1/\varepsilon$ will hold for all $c \geq c_3$ where

$$c_3 \triangleq \begin{cases} 1, & \alpha \geq \gamma, \\ \max(1, \beta/\alpha), & \alpha < \gamma, f(\beta/\alpha) \geq 0, \\ \text{the unique root of } f(c) = 0 \text{ in } (\max(\beta/\alpha, 1), \infty), & \text{otherwise.} \end{cases} \quad (6.9)$$

Further, $c_3$ is bounded as

$$\frac{\gamma}{\alpha} \leq c_3 \leq \frac{\gamma}{\alpha} + \begin{cases} 2\frac{\beta}{\alpha} \ln \frac{\beta+\gamma}{\alpha}, & \alpha \leq \beta + \gamma, \\ \ln \frac{\gamma}{\alpha}, & \text{otherwise.} \end{cases} \quad (6.10)$$

We have now established a main result of the paper:

**Theorem 6.1.** *Let $x^* > 1/2$ be the solution of the* MAXGRENT *problem* (2.7) *with $y = \mu$. In the $x^*$-case, further assume that $\|x^*\|_\infty < 2\|x^*\|_1$. Let $c_1, c_2$ denote the right-hand sides of the inequalities* (6.5) *in the $G^*$-case and* (6.7) *in the $x^*$-case. When the problem is scaled by a factor $c$,*

1. *If $c \geq \max(c_1, c_2)$, the count vector $\nu^* = [cx^*]$ is in the set $A_{n^*}(\delta, \eta)$. If, in addition, $\delta, \vartheta$ satisfy* (6.6) *then $\nu^*$ is in $A_{n^*}(\delta, \vartheta)$.*

2. *The concentration threshold is $\hat{c} \triangleq \max(c_1, c_2, c_3)$ where $c_3$ is as in* (6.9) *and* (6.10). *Then for any $c \geq \hat{c}$, in the $G^*$-case of §4 we have*

$$\frac{\#_\mu \nu^*}{\#_\mu \mathcal{B}_{n_1:n_2}(\delta, \eta)} \geq \frac{1}{\varepsilon} \quad \text{and} \quad \frac{\#_\mu A_{n_1:n_2}(\delta, \eta)}{\#_\mu (N_{n_1:n_2} \cap C(\delta))} \geq 1 - \varepsilon,$$

*and in the $x^*$-case of §5, if $\delta, \vartheta$ satisfy* (6.6),

$$\frac{\#_\mu \nu^*}{\#_\mu \mathcal{B}_{n_1:n_2}(\delta, \vartheta)} \geq \frac{1}{\varepsilon} \quad \text{and} \quad \frac{\#_\mu A_{n_1:n_2}(1/2\rho_\infty, \vartheta)}{\#_\mu (N_{n_1:n_2} \cap C(\delta))} \geq 1 - \varepsilon.$$

One aspect of the uncertainty in the problem (2.7) is the width of the interval $[s_1, s_2]$ from (2.10). It can be seen from the expression for $K$ that the scaling factor required to achieve the degree of concentration specified by $\varepsilon$ behaves intuitively with this width. (See also Example 4.2 in [Oik17].) This behavior is already apparent from Lemmas 4.2 and 5.4.

Finally, when the concentration is with respect to $\ell_\infty$ distance from $x^*$, at the expense of introducing some additional complexity, it is possible to improve the constant $K$ by using a lower bound on the number of lattice points in $A_n(\delta, \vartheta)$, as was done in §III.D of [OG16].

# 7 Probabilistic formulation

In §4 and §5 we took $\mu$ to be an $m$-vector either $\geq 1$ or of positive integers. What happens if we take it to be a density vector? Then the interpretation is that we have a 1-dimensional array of $m$ bins and we assign indistinguishable balls to them one-by-one, so that ball $i$ goes to bin $j$ with probability $\mu_j$, independent of $i$. Given $\nu = (\nu_1, ..., \nu_m)$ with sum $n$, (2.1) then becomes the total probability $\Pr_\mu(\nu)$ of all $n$-sequences of assignments (bin numbers) that have count vector $\nu$. As before, the number $n$ of balls we assign is unspecified, subject only to $n_1 \leq n \leq n_2$.

On the other hand, it is also possible to keep the purely discrete, combinatorial interpretation of Fig. 2.1: $\mu$ remains a vector of integers and we divide both sides of (2.1) by $r^n$; the result is the same as using the rational density vector $\mu/r$.

Below we outline how, with this new interpretation of $\mu$, the developments in §4 and §5 carry through, so that we end up having concentration in the usual probabilistic (e.g. [DP09]) sense: *under the measure $\mu$, most of the probability is concentrated around the single vector $\nu^*$; this vector has exponentially higher probability than the entire set $\mathcal{B}_{n_1:n_2}$.*

## 7.1 Lower bounds

When $\mu$ is a density, as opposed to a count, vector the main difference is that by Proposition 2.2 the relative entropy $G(\nu \| \mu)$ is always $\leqslant 0$, and so $G^*(\mu) \leqslant 0$ also[7.1]. Therefore in §4 we have $(1+\eta) G^*(\mu)$ instead of $(1-\eta) G^*(\mu)$ in the definitions (4.1) and (4.3) of the sets $\mathcal{A}_n, \mathcal{B}_n$ and $\mathcal{A}_{n_1:n_2}$, $\mathcal{B}_{n_1:n_2}$. Next, when $\#_\mu \nu$ becomes $\Pr_\mu(\nu)$, (4.5) turns into

$$e^{-m/12} S(\nu) e^{G(\nu\|\mu)} \leqslant \Pr_\mu(\nu) \leqslant S(\nu) e^{G(\nu\|\mu)}.$$

The development that leads to (4.7) still holds, but in (4.9) we need to substitute $1/\mu_i$ for $\mu_i$. Then the factor $1/\prod_i \sqrt{\mu_i}$ in the $C_0(x^*, \mu)$ of (4.10) becomes $\prod_i \sqrt{\mu_i}$, so that finally the probabilistic analogue of Lemma 4.2 is that if $x^* > 1/2$, then

$$\frac{\Pr_\mu(\nu^*)}{\Pr_\mu(\mathcal{B}_{n_1:n_2}(\delta,\eta))} > \frac{(m+1)\Gamma(m/2) e^{-m/12}}{2\pi^{m/2}} \prod_{i=1}^m \sqrt{\mu_i} \frac{C_2(x^*) C_4(x^*)}{C_3(s_1,s_2)} e^{-\eta G^*(\mu)}. \tag{7.1}$$

The other consequence of the change in (4.9) is that in the second claim of Lemma 4.2, the $\sum_{1 \leqslant i \leqslant m} \ln \frac{\mu_i}{\chi_i^*}$ in the condition on $\eta$ becomes $\sum_{1 \leqslant i \leqslant m} \ln \frac{1}{\mu_i \chi_i^*}$.

Turning to the development in §5, the same change as above is needed in the constant $C_0(x^*, \mu)$, and Lemma 5.2 has no dependence on whether $G(x \| \mu) \geqslant 0$ or not, so the probabilistic analogue of Lemma 5.4 is that if $x^* > 1/2$ and $\|x^*\|_\infty < 2\|x^*\|_1$, then

$$\frac{\Pr_\mu(\nu^*)}{\Pr_\mu(\mathcal{B}_{n_1:n_2}(\delta,\vartheta))} \geqslant \frac{(m+1)\Gamma(m/2) e^{-m/12}}{2\pi^{m/2}} \prod_{i=1}^m \sqrt{\mu_i} \frac{C_2(x^*) C_4(x^*)}{C_3(s_1,s_2)} e^{\vartheta^2 K(\vartheta,x^*)\|x^*\|_\infty - \delta \tilde{G}}. \tag{7.2}$$

## 7.2 Concentration threshold

Using the bounds (7.1) and (7.2) the development of §6 carries through until we get to (6.8). There we have a constant $K'$, obtained by putting $\sqrt{\mu_1 \cdots \mu_m}$ in place of $1/\sqrt{\mu_1 \cdots \mu_m}$ in $K$, and a $\gamma'$ which is $\gamma$ with $K'$ in place of $K$. Further, we see from the proof of (6.9) that $K' \ll 1$ holds a fortiori, so we still have $\gamma' > 0$ in (6.8).

Thus we finally obtain a probabilistic analogue of Theorem 6.1:

**Theorem 7.1.** *Let $\mu$ be a probability vector, and let $x^*$ be a solution of the* MAXGRENT *problem* (2.7) *with $y = \mu$, satisfying the same conditions as in Theorem 6.1. Let $c_1, c_2$ be as in Theorem 6.1, but with $\mu_i \mapsto 1/\mu_i$ in the $G^*$-case. Then when the problem is scaled by the factor $c \geqslant 1$,*

1. *If $c \geqslant \max(c_1, c_2)$, the count vector $\nu^* = [cx^*]$ is in the set $\mathcal{A}_{n^*}(\delta, \eta)$. If, in addition, $\delta, \vartheta$ satisfy (6.6) then $\nu^*$ is in $\mathcal{A}_{n^*}(\delta, \vartheta)$. Further, $\nu^*$ is such that*

$$\left. \begin{array}{c} \dfrac{\Pr_\mu(\nu^*)}{\Pr_\mu(\mathcal{B}_{n_1:n_2}(\delta,\eta))} \\[1em] \dfrac{\Pr_\mu(\nu^*)}{\Pr_\mu(\mathcal{B}_{n_1:n_2}(\delta,\vartheta))} \end{array} \right\} \geqslant K' \frac{e^{\alpha c}}{c^\beta},$$

---

[7.1]. Recall Proposition 2.5(2).

where K′ is the K of (6.8) but with $\mu_i \mapsto 1/\mu_i$, and $\alpha, \beta$ are as in (6.8).

2. Let $f(c)$ be as in Theorem 6.1, but with $\gamma$ modified by $\mu_i \mapsto 1/\mu_i$, and let $c_3$ be as defined there; then $c_3$ is bounded as in (6.10).

3. The concentration threshold is $\hat{c} \triangleq \max(c_1, c_2, c_3)$. If $c \geqslant \hat{c}$, we have in the $G^*$-case

$$\frac{\Pr_\mu(\nu^*)}{\Pr_\mu(\mathcal{B}_{n_1:n_2}(\delta,\eta))} \geqslant \frac{1}{\varepsilon} \quad \text{and} \quad \frac{\Pr_\mu(\mathcal{A}_{n_1:n_2}(\delta,\eta))}{\Pr_\mu(\mathcal{N}_{n_1:n_2} \cap C(\delta))} \geqslant 1-\varepsilon,$$

and in the $x^*$-case

$$\frac{\Pr_\mu(\nu^*)}{\Pr_\mu(\mathcal{B}_{n_1:n_2}(\delta,\vartheta))} \geqslant \frac{1}{\varepsilon} \quad \text{and} \quad \frac{\Pr_\mu(\mathcal{A}_{n_1:n_2}(\delta,\vartheta))}{\Pr_\mu(\mathcal{N}_{n_1:n_2} \cap C(\delta))} \geqslant 1-\varepsilon.$$

There are some other differences with the results of §6 that may not be apparent from the statement of the theorem. As mentioned in §7.1, the sets $\mathcal{A}, \mathcal{B}$ are defined slightly differently here than in the $G^*$-case of §4. Second, in §2.3, the condition (2.10) that the sum of $x$ is bounded was necessary for the *existence of a maximum*. Such a condition is not needed here because $G(x \| \mu)$ is bounded from above by 0; but it *is* still needed for concentration to occur. Third, we can no longer guarantee that $x^*$ is the unique maximizer of $G(x \| \mu)$; recall Lemma 2.6 and Proposition 2.5(2).

# 8 Examples

We give an example, adapted from [Oik17], from the field of transportation analysis/planning. Classic MAXENT has beeen frequently applied to problems in this area (see [KK92]), but our example is of interest because it is not treatable by MAXENT. While it does not include the complexities of a real problem, the example illustrates fundamental issues that are sometimes obscured by complexity.

All results in this section were generated by code in the JULIA language, https://julialang.org, using its JUMP optimization framework, [DHL17].

**Example 8.1.** Fig. 8.1 shows four cities connected by road segments. We assume that vehicles travelling between cities follow the most direct route, and that there is no traffic from a city to itself.

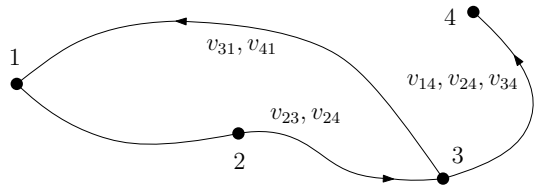

**Figure 8.1.** Four cities connected by bidirectional road segments. Some segments have been labelled with an arrow indicating the direction for which we have traffic information, and with the flows that go through them in that direction.

The number of vehicles in each city is known, which puts upper bounds on the number that leave each city; also we have either a lower bound or exact data for the number of vehicles on the road segments $2\to 3$, $3\to 1$, and $3\to 4$. From this information we want to infer how many vehicles travel from city $i$ to city $j$, i.e. the matrix of counts $[v_{ij}]_{4\times 4}$ with diagonal equal to 0. So suppose the information on $v$ is

$$v_{ii}=0, \quad \sum_j v_{ij} \leqslant 200, 240, 160, 180, \quad i=1,\dots,4 \tag{8.1}$$
$$v_{23}+v_{24}\geqslant 154, \quad v_{31}+v_{41}=122, \quad v_{14}+v_{24}+v_{34}=172,$$

where the last three constraints reflect the "direct route" assumption. The 12-element vector $x$ for problem (2.7) is $(v_{12},v_{13},v_{14},v_{21},v_{23},v_{24},\dots,v_{43})$, and we have $s_1=294, s_2=780$, and $\rho_\infty=122$. [Note that if we knew that *all* vehicles in a city leave the city, then we could define a *frequency* matrix by dividing $v$ by $200+\cdots+180$ and thus formulate a MAXENT problem.] The prior is $\mu=(2,2,1,\dots,1,2)$, expressing some preference for short trips out of city 1 and city 4.

Table 8.1 shows the MAXGRENT integral solution and also the MINIDIV integral solution.

$$v^* = \begin{bmatrix} 0 & 79 & 79 & 41 \\ 79 & 0 & 79 & 82 \\ 65 & 47 & 0 & 49 \\ 57 & 41 & 82 & 0 \end{bmatrix}, \quad \hat{v} = \begin{bmatrix} 0 & 2 & 2 & 14 \\ 1 & 0 & 10 & 144 \\ 61 & 1 & 0 & 14 \\ 61 & 1 & 2 & 0 \end{bmatrix}.$$

**Table 8.1.** The MAXGRENT solution $v^*$ and MINIDIV solution $\hat{v}$, both with $\mu=(2,2,1,\dots,1,2)$. We have $s^*=s_2=780$, $G^*=2079.4$, $G(\hat{v})=485.6$, and $\|\hat{v}-x^*\|_\infty = 79.93$.

Why is the MAXGRENT solution to be preferred? Table 8.2 allows us to compare it with the MINIDIV solution using the $G$-values. To have $\hat{v}\in \mathcal{B}_{294:780}(0.01,\eta)$ we need $\eta < 1-G(\hat{v})/G^* = 0.7669$, and even using the $\eta=0.2$ entry of the table we see that $\#_\mu v^*/\#_\mu \mathcal{B}_{294:780}(0.01,0.2) > 10^{144}$, an impressive factor, especially since the denominator contains not only the MINIDIV solution, but all other possible solutions as well[8.1]. Also, it is clear that the prior plays a very different role in the two methods, as we remarked at the end of §3.

Table 8.3 shows the factor by which the problem needs to be scaled (concentration threshold) to achieve concentration to degree $\varepsilon$ in a region within 1% of the MAXGRENT value $G^*$ and in a region within 0.5% of $G^*$. As expected, $\hat{c}$ is sensitive to $\eta$ and relatively insensitive to $\varepsilon$.

| $\delta$ | $\eta$ | $\#_\mu v^*/\#_\mu \mathcal{B}$ |
|---|---|---|
| 0.01 | 0.04 | 0.542 |
|  | 0.05 | $5.8\cdot 10^8$ |
|  | 0.06 | $6.2\cdot 10^{17}$ |
|  | 0.07 | $6.7\cdot 10^{26}$ |
|  | 0.08 | $7.2\cdot 10^{35}$ |
|  | 0.09 | $7.7\cdot 10^{44}$ |
|  | 0.10 | $8.3\cdot 10^{53}$ |
|  | 0.20 | $1.7\cdot 10^{144}$ |

**Table 8.2.** The lower bound of Lemma 4.2.

| $\delta$ | $\eta$ | $\varepsilon$ | $\hat{c}$ |
|---|---|---|---|
| 0.01 | 0.01 | $10^{-10}$ | 1 |
|  |  | $10^{-15}$ | 6.75 |
|  |  | $10^{-20}$ | 7.35 |
|  |  | $10^{-30}$ | 8.54 |
| 0.005 |  | $10^{-10}$ | 13.1 |
|  |  | $10^{-15}$ | 14.3 |
|  |  | $10^{-20}$ | 15.5 |
|  |  | $10^{-30}$ | 17.9 |

**Table 8.3.** The concentration threshold $\hat{c}$ of Theorem 6.1.

---

8.1. The ratios $\#_\mu v^*/\#_\mu \mathcal{B}$ are so large that a more appropriate measure would be logarithmic: the log-odds in [Jay03] §4.3, originally due to I.J. Good.

To compare using the distance measure, $\|\hat{v} - x^*\|_\infty = 79.93$. To have $\hat{v} \in \mathcal{B}_{294:780}(\delta, \vartheta)$ we need $\|\hat{v} - x^*\|_\infty > 780\,\vartheta$ which requires $\vartheta < 0.102$. However, the problem is too small for the bound of Lemma 5.4: with $\vartheta = 0.1, \delta = 0.0045$ it yields the useless lower bound $3.5 \cdot 10^{-41}$. Table 8.4 gives an idea of how large the problem must be.

| $\varepsilon$ | $\delta$ | $\vartheta$ | $\hat{c}$ | $\delta$ | $\vartheta$ | $\hat{c}$ | $\delta$ | $\vartheta$ | $\hat{c}$ |
|---|---|---|---|---|---|---|---|---|---|
| $10^{-3}$ | $10^{-5}$ | 0.1 | 410 | $10^{-5}$ | 0.05 | 2020 | $10^{-6}$ | 0.02 | 12500 |
| $10^{-5}$ | | | 410 | | | 2080 | | | 12900 |
| $10^{-10}$ | | | 441 | | | 2210 | | | 13600 |
| $10^{-15}$ | | | 472 | | | 2350 | | | 14400 |
| $10^{-20}$ | | | 502 | | | 2490 | | | 15100 |

**Table 8.4.** The threshold $\hat{c}$ of Theorem 6.1 for concentration around $x^*$. Note the interaction between $\delta$ and $\vartheta$.

Finally, we note that Proposition 3.1 says that as the prior becomes a large multiple of $\mu = (2, 2, 1, \ldots, 1, 2)$, the factor approaching $s^* = 780$, the MINIDIV $\hat{v}$ approaches the MAXGRENT $v^*$. In fact, even with the prior $100\,\mu$ the MINIDIV solution $u^*$ turns out to be a very good approximation to the MAXGRENT $x^*$.

Continuing Example 8.1, now consider the density or probability priors $(2,2,1,\ldots,1,2)/15$ and $(1,\ldots,1)/12$. The MAXGRENT and MINIDIV solutions are shown in Table 8.5.

$$v^* = \begin{bmatrix} 0 & 79 & 79 & 43 \\ 50 & 0 & 74 & 80 \\ 65 & 46 & 0 & 50 \\ 57 & 41 & 82 & 0 \end{bmatrix}, \begin{bmatrix} 0 & 61 & 61 & 53 \\ 61 & 0 & 82 & 72 \\ 59 & 54 & 0 & 47 \\ 63 & 58 & 58 & 0 \end{bmatrix}, \hat{v} = \begin{bmatrix} 0 & 0 & 0 & 10 \\ 0 & 0 & 1 & 153 \\ 61 & 0 & 0 & 10 \\ 61 & 0 & 0 & 0 \end{bmatrix}.$$

**Table 8.5.** The MAXGRENT solutions $v^*$ with the density version $\mu = (2,2,1,\ldots,1,2)/15$ of the prior used in Tables 8.2–8.4 and with the uniform prior $\mu = (1,\ldots,1)/12$. The first solution has $s^* = 743.6, G^* = -23.4$ and the second $s^* = 729.4, G^* = -7.07$. The integral MINIDIV solutions $\hat{v}$ are the same for both priors, and have $G(\hat{v}) = -434.5$ and $-368.4$ respectively.

A noticeable difference with the case $\mu \geqslant 1$ is that all else being equal, the concentration threshold becomes higher, as can be seen from Table 8.6. This is not surprising given (7.1) and (7.2).

| $\delta$ | $\vartheta$ | $\varepsilon$ | $\hat{c}$ |
|---|---|---|---|
| $10^{-5}$ | 0.1 | $10^{-3}$ | 414 |
| | | $10^{-5}$ | 425 |
| | | $10^{-10}$ | 455 |
| | | $10^{-15}$ | 484 |
| | | $10^{-20}$ | 513 |

**Table 8.6.** The concentration threshold $\hat{c}$ of Theorem 6.1 vs. $\varepsilon$ with the prior $\mu = (2,2,1,\ldots,1,2)/15$.

| $\delta$ | $\eta$ | $c=5$ | $c=10$ | $c=15$ |
|---|---|---|---|---|
| 0.01 | 0.4 | $7.5 \cdot 10^{-31}$ | $4.5 \cdot 10^{-14}$ | $7.7 \cdot 10^4$ |
| | 0.5 | $9.0 \cdot 10^{-26}$ | $6.4 \cdot 10^{-4}$ | $1.3 \cdot 10^{20}$ |
| | 0.6 | $1.1 \cdot 10^{-20}$ | $9.1 \cdot 10^6$ | $2.2 \cdot 10^{35}$ |
| | 1 | 2.1 | $3.6 \cdot 10^{47}$ | $1.7 \cdot 10^{96}$ |
| | 1.5 | $5 \cdot 10^{25}$ | $2 \cdot 10^{98}$ | $2.3 \cdot 10^{172}$ |

**Table 8.7.** The lower bound of (7.1) on $\Pr_\mu(v^*)/\Pr_\mu(\mathcal{B})$ with $\mu = (2,2,1,\ldots,1,2)/15$ and some scalings $c$ of the problem.

Table 8.7 shows that the problem is not large enough for the lower bound of Lemma 4.2 to be meaningful, unless we apply some scaling by $c$ and $\eta$ is rather large. Further, we see from the $\eta = 1.5$ row that the ratio $G(\hat{v})/G^* = 18.9$ renders the MINIDIV solution exceedingly improbable.

We also note that for MAXGRENT the probability prior can be interpreted in either of the two ways mentioned at the beginning of §7. But in the case of MINIDIV its interpretation is not clear. While it might be argued that a prior which is a density vector is inappropriate for MINIDIV in this problem, whether the prior is a density or not is not an issue for MAXGRENT.

## 9 Conclusion

We presented and studied a new generalization of relative entropy to non-negative real vectors which do not necessarily sum to 1. Its combinatorial interpretation is in terms of count vectors, whereas that of ordinary entropy and relative entropy is in terms of frequency vectors. We believe that our results, especially the concentration property, provide support for an extension of the well-known MAXENT framework for inference of probability distributions to what we have called here the MAXGRENT method, itself an extension of the MAXGENT of [Oik17], for inference of count vectors. Still, there are many more aspects of MAXGRENT and its relationships with other methods that remain to be explored.

## Appendix A  Proofs

**Proof of Proposition 2.2**

1. This can be seen from (2.2) and the second line of (2.4), since $G(x)$ is always non-negative.
2. From the log-sum inequality, $\sum_i x_i \ln(x_i/y_i) \geqslant (\sum_i x_i) \ln(\sum_i x_i / \sum_i y_i) = (\sum_i x_i) \ln(\sum_i x_i)$. Then $G(x \| y) \leqslant 0$ follows from (2.3). Another proof uses (2.5): we have $r = 1$, and the divergence $D(\chi \| \psi)$ is always non-negative.
3. The value of $G(y \| y)$ follows from (2.5).
4. If $y \geqslant 1$, all partial derivatives of $G(x \| y)$ are $\geqslant 0$ at $x$; and if $y_i > 1$, $\partial G / \partial x_i > 0$. Thus $x'$ can be reached from $x$ by a sequence of changes that cannot decrease, and may increase, the value of $G$. More formal proof: by the mean value theorem there is a point $u$ on the line segment (convex set) between $x$ and $x'$ s.t. $G(x' \| y) - G(x \| y) = \nabla_x G(u \| y) \cdot (x' - x)$. The elements of the gradient at $u$ are $\ln\left((y_i/u_i) \sum_j u_j\right)$, and we already have $x' - x \geqslant 0$. A simple sufficient condition for $\nabla_x G(u \| y)$ to be $\geqslant 0$ is $y \geqslant 1$. [A weaker sufficient condition is that $\bar{u} \leqslant y$, where $\bar{u}$ is the normalized $u$.]
5. It is pointed out in [Oik17] that the generalized entropy (2.2) is such that $G(\alpha x) = \alpha G(x)$ for any $\alpha > 0$. That the generalized relative entropy $G(x \| y)$ has the same property then follows from (2.4).
6. By the second line of (2.4), for fixed $y$, $G(x \| y)$ is the sum of the function $G(x)$ of (2.2), shown concave in §2.2 of [Oik17], and of a linear (affine) function; hence it is concave. That it is not strictly concave follows from item 5 above. Also, $G((x_1, x_2) \| (1,1))$ is an example. Since strong concavity implies strict concavity, $G$ can't be strongly concave either.
7. By concavity, $\forall \alpha, \beta > 0$, $G\left(\frac{\alpha}{\alpha+\beta} x + \frac{\beta}{\alpha+\beta} y \,\big\|\, z\right) \geqslant \frac{\alpha}{\alpha+\beta} G(x \| z) + \frac{\beta}{\alpha+\beta} G(y \| z)$. Then by the positive homogeneity property in Proposition 2.2(5), multiplying both sides by $\alpha + \beta$ yields the desired result.

8. The lower bound follows simply from the 2nd line of (2.4) by noting that the generalized entropy $G(x)$ of (2.2) is always $\geq 0$, and that $\sum_i x_i \ln y_i \geq (\min_i \ln y_i) \sum_i x_i$. The upper bound follows from the definition (2.3) by noting that by the log-sum inequality $\sum_i x_i \ln (x_i/y_i) \geq (\sum_i x_i) \ln (\sum_i x_i / \sum_i y_i)$.

9. Suppose we group the $i$th and $j$th elements of $x$ as well as those of $y$: the value of $G(x\|y)$ is unaffected when the same permutation is applied to both arguments, so after a suitable permutation we can act as if we are grouping the first two elements of each. Then from (2.3)

$$G((x_1+x_2, x_3, \ldots, x_m) \| (y_1+y_2, y_3, \ldots, y_m)) \geq G(x\|y) \quad \Leftrightarrow$$
$$(x_1+x_2) \ln \frac{x_1+x_2}{y_1+y_2} \leq x_1 \ln \frac{x_1}{y_1} + x_2 \ln \frac{x_2}{y_2},$$

which holds by virtue of the log-sum inequality.

**Proof of Proposition 2.5**

1. This follows from Proposition 2.2(4).

2. By Proposition 2.2(2), $G(x\|y) = -sD(\chi\|\psi) \leq 0$, and $D(\chi\|\psi)$ becomes 0 iff $\chi = \psi$.

3. First recall the upper bound in Proposition 2.2(7). With only the constraint $\sum_i x_i \leq s$, the expression for $x^*$ in (2.8) reduces to $x_j^* = y_j (\sum_i x_i^*) e^{-\xi^*}$, $\xi^* (\sum_i x_i^* = s)$, $\xi^* \geq 0$. First assume that $\xi^* > 0$. Then we must have $\sum_i x_i^* = s$ and summing the expression for $x_j^*$ over all $j$ we find that $e^{-\lambda^*} r = 1$. So $x_j^* = (s/r) y_j$ as claimed; the result for the equality constraint follows from that for the inequality constraint. Next, using the above $x^*$ in (2.3) it follows that $G(x^*\|y) = s \ln r$ as claimed.

   Now assume that $\xi^* = 0$; we then see that we must have $r = 1$, i.e. $y$ must be a density vector. In this case, *any* feasible $x$, i.e. with $\sum_i x_i \leq s$, maximizes $G$. The value of the maximum is $s \ln r$, as for $\xi^* > 0$, and this value is 0. As a simple example of this situation consider maximizing $G((x_1, x_2) \| (0.5, 0.5))$ subject to $x_1 + x_2 \leq s$.

4. The constraint $\sum_i x_i \leq s_2$ is always implicitly present. By part 3, if that were the only constraint, we would have $G^*(y) = s_2 \ln r$. In the presence of additional constraints $G^*(y) \leq s_2 \ln r$. Now assume the sum $s^*$ of the solution $x^*$ is known; by part 3, solving the problem under only the constraint $\sum_i x_i = s^*$ results in $G^*(y) = s^* \ln r$. When we add the rest of the constraints the sum remains at $s^*$ and the value of $G^*(y)$ can only decrease or remain the same.

**Proof of Lemma 2.6**

The proof uses only properties of $G(x\|y)$, and relies only on that $C$ is convex, closed, and bounded.

1. The proof of the uniqueness of the maximum is by contradiction. Assume that $u, v$ are two (distinct) global maximizers of $G$ over $C$. It is not possible that both of them have the same sum $s$: under the condition $\sum_i x_i = s$, we have $G(x\|y) = s \ln s - \sum_i x_i \ln x_i + \sum_i x_i \ln y_i$ by (2.3). But this is a strictly concave function of $x$ (sum of the strictly concave Shannon entropy[A.1] and of a linear function), so it has a unique global maximizer over the convex domain $C \cap \{x | \sum_i x_i = s\}$.

---

A.1. Extended to all $x \geq 0$. This is strictly concave, the Hessian is easily seen to be negative definite.

Next let $u$ and $v$ have different sums. We will derive a condition *necessary* for both $u$ and $v$ to maximize $G$ and show that it is contradicted by the positive homogeneity of $G$. Under our assumption of $G(u\|y) = G(v\|y) = G^*$, the concavity of $G$ implies that any point on the line segment between $u$ and $v$ must yield a value $\geqslant G^*$. Since $G$ never exceeds $G^*$, this value must be $G^*$. Thus the function

$$f(\alpha) = G(\alpha u + (1-\alpha) v \| y), \quad \alpha \in [0,1]$$

must be a constant. Therefore $f'(\alpha)$ must be 0 for all $\alpha \in (0,1)$. Rather than $f'(\alpha)$, it is easier to deal with the expression for $f''(\alpha)$, which also has the advantage of not involving $y$. The constancy of $f'(\alpha)$ implies that we must have $f''(\alpha) \equiv 0$:

$$f''(\alpha) = \frac{(\sum_i u_i - \sum_i v_i)^2}{\alpha \sum_i u_i + (1-\alpha) \sum_i v_i} - \sum_i \frac{(u_i - v_i)^2}{\alpha u_i + (1-\alpha) v_i}.$$

We will consider the condition $f''(1/2) = 0$, and set $u_i - v_i = z_i$, $u_i + v_i = w_i$. Then we have

$$f''(1/2) = \frac{(\sum_i z_i)^2}{\sum_i w_i} - \sum_i \frac{z_i^2}{w_i} = 0.$$

Further setting $q_i = w_i / \sum_i w_i$, since $\sum_i w_i > 0$ the above condition is equivalent to $(\sum_i z_i)^2 - \sum_i (z_i^2 / q_i) = 0$. But the l.h.s. is a strictly concave function of $q$, hence over the convex set $q > 0$, $\sum_i q_i = 1$ it attains its global maximum of 0 at a *unique* point $\hat{q}$, where $\hat{q}_i = z_i / \sum_j z_j$.

So we have shown that $f''(1/2) = 0 \Rightarrow w_i / \sum_j w_j = z_i / \sum_j z_j$ for all $i$. This is equivalent to

$$\forall i, \quad \frac{u_i + v_i}{\sum_j u_j + \sum_j v_j} = \frac{u_i - v_i}{\sum_j u_j - \sum_j v_j} \quad \text{or} \quad \frac{u_i}{v_i} = \frac{\sum_j u_j}{\sum_j v_j}. \tag{A.1}$$

This condition is *necessary* for $f'(\alpha)$ to be constant, in particular 0, hence for $f(\alpha)$ to be constant. Finally, we can assume w.l.o.g. that the points $u$ and $v$ are such that $\sum_i u_i > \sum_i v_i$, and then (A.1) implies that there is some $\alpha > 1$ s.t. $u = \alpha v$. But then by Proposition 2.2(5), $G(u\|y) = \alpha G(v\|y) > G(v\|y)$, contradicting our initial assumption that both $u$ and $v$ maximize $G$. [The last inequality holds because $G(v\|y) > 0$ when $y \geqslant 1$; recall Remark 2.4.]

2. When $y$ is a density vector the argument we gave above still holds, until we get to the inequality $\alpha G(v\|y) < G(v\|y)$, which is now reversed because $G(v\|y) \leqslant 0$; this inequality fails when $G(v\|y) = G^* = 0$.

3. If $C$ is not full-dimensional, i.e. its dimension is $< m$, then the claim is trivially true: every point of $C$ is a boundary point in (relative to) $\mathbb{R}^m$. Now suppose $C$ is full-dimensional, and $x^* \in \text{int } C$. Then for some $\eta > 0$ there exists an $\ell_\infty$-ball $\mathbb{B}(x^*, \eta)$ contained in $C$. Hence by Proposition 2.2(4) the point of $C$ obtained by adding $\eta$ to each element of $x^*$ yields a value of $G$ larger than $G^*$, a contradiction. [The polytope defined by (2.7) is full-dimensional if there are no equality constraints and also no equalities *implied* by the inequalities; see [Sch86], §8.2 for details.]

Concerning the relative boundary, on the set $C = \{x \in \mathbb{R}_+^2 \mid x_1 + x_2 = s\}$ whose interior in $\mathbb{R}^2$ is empty and whose relative boundary (w.r.t. its affine hull) consists of the two points $(0, s)$ and $(s, 0)$, $G(x \| (1,1))$ is maximized at the *interior* point $(s/2, s/2)$.

**Proof of Proposition 2.8**

We can write (2.12) as $F(x\|\mu) = F(x) - \sum_i x_i \ln \mu_i$, where $F(x) = \sum_i x_i \ln x_i - (\sum_i x_i) \ln(\sum_i x_i)$. $F(x)$ is the negative of the generalized entropy $G(x)$ introduced in [Oik17]. The function $F$ also appears in 3.51 of [RW09], and it is shown in 11.12 there that it is the *support* function of the convex set $\{u \in \mathbb{R}^m | \ln(e^{u_1} + \cdots + e^{u_m}) \leqslant 0\}$, equivalently of $\{u \in \mathbb{R}^m | e^{u_1} + \cdots + e^{u_m} \leqslant 1\}$. Now by 11.4 of [RW09], the conjugate of the support function of a closed convex set $C$ is the *indicator* function $\delta_C$ of this set; it follows that the conjugate $F^\dagger(u)$ of $F(x)$ is the indicator function of the set $\{u \in \mathbb{R}^m | \sum_i e^{u_i} \leqslant 1\}$.

Finally, by using 11(3) of [RW09], i.e. the correspondence $\varphi(x) - \alpha \cdot x \leftrightarrow_\dagger \psi(u + \alpha)$ that holds when $\varphi$ is proper, lower-semicontinuous, and convex[A.2], which $F(x)$ is, we see that the conjugate $F^\dagger(u\|\mu)$ of $F(x\|\mu)$ is the indicator function of $\{u \in \mathbb{R}^m | \sum_i e^{u_i + \ln \mu_i} \leqslant 1\}$, as claimed.

**Proof of Proposition 2.9**

The first three inequalities follow from the basic property of rounding, $\forall y \in \mathbb{R}, |[y] - y| \leqslant \tfrac{1}{2}$. For the last inequality, omitting the stars for simplicity, we have

$$\left\|\frac{\nu}{n} - \frac{x}{s}\right\|_1 = \left\|\frac{\nu}{n} + \frac{x}{n} - \frac{x}{n} - \frac{x}{s}\right\|_1 \leqslant \frac{1}{n}\|\nu - x\|_1 + \left|\frac{1}{n} - \frac{1}{s}\right|\|x\|_1 \leqslant \frac{m}{2n} + \frac{|n-s|}{ns}s \leqslant \frac{m}{2n} + \frac{m}{2n}.$$

**Proof of Proposition 2.7**

Suppose that the vectors $x^*, \lambda^*, \xi^*$ satisfy condition (2.8); we show that if $b \mapsto \alpha b$, then $\alpha x^*, \lambda^*, \xi^*$ also satisfy this condition. This means that if $x^*$ solves problem (2.7), the scaled vector $\alpha x^*$ solves the problem after the data is scaled by $b \mapsto \alpha b$. That $\alpha x^*, \lambda^*, \xi^*$ satisfy (2.8) with $b \mapsto \alpha b$ is clear from the linearity of the constraints on $x$ in the first line, and the fact that the expression for $x^*$ in the second line is invariant under $x^* \mapsto \alpha x^*$, as is the condition on $\xi^*$ in the third line.

That $G^*$ scales as claimed follows from Proposition 2.2(5). Finally, the fact that the bounds on $\sum_i x_i$ scale with $\alpha$ is just a property of linear programs in general: if $z$ solves the LP $\min_{x \in \mathbb{R}^m} \sum_i a_i x_i$ subject to $Cx \leqslant d$, then $\alpha z$ solves it with the scaled constraint $Cx \leqslant \alpha d$. Similarly for the maximum.

**Proof of Proposition 3.1**

Consider the two concave programs $\max_{x \in C(0)} G(x\|y)$ and $\max_{u \in C(0)} -\mathcal{D}(u\|v)$, with $C(0)$ as in (2.7). The Lagrangians are

$$\begin{aligned} L^G(x, \lambda, \xi) &= G(x\|y) - \lambda \cdot (Ax - b) - \xi \cdot (Cx - d), \\ L^\mathcal{D}(u, \rho, \sigma) &= -\mathcal{D}(u\|v) - \rho \cdot (Au - b) - \sigma \cdot (Cu - d), \end{aligned}$$

and the necessary and sufficient conditions for $x^*(y), u^*(v)$ to be the solutions of the concave programs are (2.8) and (3.2).

1. Given the solution $x^*(y)$ of the MAXGRENT problem, set $v = \tilde{v} = y \sum_j x_j^*$, and $\rho = \lambda^*, \sigma = \xi^*$ in (3.2). Then we see that the resulting expressions are satisfied by $(x^*(y), \lambda^*, \xi^*)$, hence $u^*(\tilde{v}) = x^*(y)$ as claimed.

2. The Lagrangian is now $L^G(x, \lambda, \xi, \lambda_0)$, containing the extra term $-\lambda_0 \sum_i u^*(v)$ and the solution $(x^*, \lambda^*, \xi^*, \lambda_0^*)$ is such that

$$x_j^* = y_j(\textstyle\sum_i x_i^*) e^{-(\lambda^* \cdot A_{\cdot j} + \xi^* \cdot C_{\cdot j} + \lambda_0^*)}, \quad \xi_j^*(C_{\cdot j} x^* - d_j) = 0, \quad \xi_j^* \geqslant 0, \quad 1 \leqslant j \leqslant m. \tag{A.2}$$

---

A.2. [RW09] use * to denote conjugacy, but we've used that to denote optimality, so we denote it by † instead.

Given the MINIDIV solution $u^*(v)$, set in the above expressions $y = \tilde{y} = v/\sum_i u^*(v)$, $\sum_i x_i^* = \sum_i u_i^*, \lambda^* = \rho^*, \xi^* = \sigma^*, \lambda_0^* = 0$. Then $(u^*(v), \rho^*, \sigma^*, \lambda_0^*)$ satisfies (A.2) and solves the MAXGRENT problem, as claimed.

**Proof of Proposition 4.1**

Let $\zeta$ be an $m$-vector all of whose components equal $\zeta$. In the hypercube $\|y - x\|_\infty \leqslant \zeta$, $G(y\|\mu)$ attains its minimum at $x - \zeta$ and its maximum at $x + \zeta$, so we have $G(x - \zeta\|\mu) \leqslant G(y\|\mu) \leqslant G(x + \zeta\|\mu)$. The lower bound of the lemma is based on Lemma 2.1 of [Oik17] which bounds $G(x - \zeta)$ from below; the proof is the same as the one given there with the simple modification $G(y\|\mu) = G(y) + \sum_i y_i \ln \mu_i$. The upper bound of the lemma is a bound on $G(x + \zeta\|\mu)$ from above, and follows likewise by changing $\zeta$ to $-\zeta$.

**Proof of Proposition 5.1**

The proof relies only on the convexity of $C$, and not on the fact that it is polyhedral. Let $\|\cdot\|$ be any $\ell_p$ norm.

1. For $a \in (0,1)$, let $y = ax^* + (1-a)x$ be a point on the line segment between $x$ and $x^*$. Since all of the segment is in $C$ because $C$ is convex, $y \in C$. Then by the concavity of $G$,

$$G(y\|\mu) \geqslant aG(x^*\|\mu) + (1-a)G(x\|\mu) > aG(x\|\mu) + (1-a)G(x\|\mu) = G(x\|\mu),$$

where the strict inequality follows from $x \neq x^*$ and the fact that $x^*$ is the unique maximizer of $G(\cdot\|\mu)$ over $C$. So $G(y\|\mu) > G(x\|\mu)$ for any $a \in (0,1)$.

Now let $x$ be outside $K^*(\vartheta^*)$, and choose $\alpha$ so that $y$ is on the surface of $K^*(\vartheta)$: $\|x^* - y\| = \vartheta \|x^*\| \Leftrightarrow (1-a)\|x - x^*\| = \vartheta \|x^*\|$. So the desired $\alpha$ is $1 - \vartheta \frac{\|x^*\|}{\|x - x^*\|}$, and it is in $(0,1)$ because $\|x - x^*\| > \vartheta \|x^*\|$ for $x$ outside $K^*(\vartheta)$.

2. As above, with $y = ax^* + (1-a)x$ we have $G(y\|\mu) > G(x\|\mu)$. Then $|\sum_i y_i - s^*| = (1-a)|\sum_i x_i - s^*|$, so $|\sum_i y_i - s^*| \leqslant \vartheta \|x^*\|$ iff $a \geqslant 1 - \vartheta \|x^*\|/|\sum_i x_i - s^*|$. Since $x \notin D^*(\vartheta)$, $a \in (0,1)$.

**Proof of Proposition 5.3**

We prove both parts of the proposition by means of one example. Let $m = 2$ and let $C$ be the very simple set $\{x \in \mathbb{R}_+^2 \mid x_1 + x_2 = s\}$. By Proposition 2.5, $x^* = \left(\frac{s\mu_1}{\mu_1 + \mu_2}, \frac{s\mu_2}{\mu_1 + \mu_2}\right)$, $s^* = s$. For some $\zeta > 0$ consider $z = x^* + (\zeta, -\zeta)$, a point to the right and below $x^*$ on the line $x_1 + x_2 = s$. By the mean value theorem there is a point $y$ between $z$ and $x^*$ on this line s.t. $G(x^*\|\mu) - G(z\|\mu) = \nabla_x G(y\|\mu) \cdot (x^* - z)$. Since $\nabla_x G(y\|\mu) = \left(\ln \frac{\mu_1 s}{y_1}, \ln \frac{\mu_2 s}{y_2}\right)$ and $\|z - x^*\|_\infty = \zeta$, it follows that

$$\frac{G(x^*\|\mu) - G(z\|\mu)}{\|z - x^*\|_\infty} = \ln\left(\frac{\mu_2}{\mu_1}\frac{y_1}{y_2}\right) < \ln\left(\frac{\mu_2}{\mu_1}\frac{x_1^* + \zeta}{x_2^* - \zeta}\right), \tag{A.3}$$

because $y_1 < z_1$ and $y_2 > z_2$ imply $y_1/y_2 < z_1/z_2$.

1. Now let $\zeta \leqslant \tilde{\zeta}$, so $\|z - x^*\| \leqslant \tilde{\zeta}$ as assumed in part 1. Noting that $\mu_2 x_1^* = \mu_1 x_2^* = \frac{\mu_1 \mu_2 s}{\mu_1 + \mu_2}$, the claim follows for the $\ell_\infty$ norm by choosing $\zeta > 0$ small enough to make the argument of the log $\leqslant 1 + \varepsilon$ (easily seen for $\mu_1 = \mu_2 = 1$). Next, for the $\ell_1$ norm we have $\|z - x^*\|_1 = 2\zeta$ so (A.3) becomes

$$\frac{G(x^*\|\mu) - G(x\|\mu)}{\|x - x^*\|_1} = \frac{1}{2}\ln\left(\frac{\mu_2}{\mu_1}\frac{y_1}{y_2}\right) < \frac{1}{2}\ln\left(\frac{\mu_2}{\mu_1}\frac{x_1^* + \zeta}{x_2^* - \zeta}\right)$$

as above. For the $\ell_2$ norm we have $\|z - x^*\|_2 = \sqrt{2}\zeta$, so analogously, and the claim is established for these norms also.

2. We see from (A.3) that a *necessary* condition for the statement of part 2 to hold at $z = x^* + (\zeta, -\zeta)$, where now we just assume $\zeta > 0$, is that

$$\zeta \geq \vartheta \|x^*\| \quad \Rightarrow \quad \zeta \ln\left(\frac{\mu_2}{\mu_1} \frac{x_1^* + \zeta}{x_2^* - \zeta}\right) > \kappa \vartheta \|x^*\|.$$

If we take $\mu_1 = \mu_2$ and $\zeta = \vartheta \|x^*\|$, the r.h.s. above reduces to

$$\ln \frac{s^*/2 + \vartheta \|x^*\|}{s^*/2 - \vartheta \|x^*\|} > \kappa. \tag{A.4}$$

With the $\ell_\infty$ norm we have $\|x^*\|_\infty = s^*/2$ so (A.4) becomes $\ln \frac{1+\vartheta}{1-\vartheta} > \kappa$. Therefore any $\kappa > 0$ satisfying it must depend on $\vartheta$, and this establishes the claim for $\|\cdot\|_\infty$. Likewise, for the $\ell_1$ and $\ell_2$ norms we have $\|x^*\|_1 = s^*$ and $\|x^*\|_2 = s^*/\sqrt{2}$, so (A.4) becomes $\ln \frac{1/2+\vartheta}{1/2-\vartheta} > \kappa$ and $\ln \frac{1/\sqrt{2}+\vartheta}{1/\sqrt{2}-\vartheta} > \kappa$ respectively.

**Proof of (6.9) and (6.10)**

1. We have already discussed $\alpha > 0$ after (6.6), so the first thing we need for (6.9) is to show that $\gamma > 0$. We do this by showing that $K \ll 1$. Noting that $s^* \chi_i^* = x_i^*$ and all $\mu_i \geq 1$,

$$K < \frac{(m+1)\Gamma(m/2) e^{-m/12}}{2\pi^{m/2}} \frac{1}{\left(\sqrt{s_2 + m/2 + 2} + \sqrt{m}\right)^{m+1} - \left(\sqrt{s_1 - m/2} + \sqrt{m}\right)^{m+1}}.$$

The second fraction above is largest when $s_1 = s_2 = s$, and the result is maximized when $s = m/2$; this is achievable by $s = s^*$ when all $x_i^* = 1/2$. So then

$$K < \frac{(m+1)\Gamma(m/2) e^{-m/12}}{2\pi^{m/2}} \frac{1}{\left(\sqrt{m+2} + \sqrt{m}\right)^{m+1} - (\sqrt{m})^{m+1}}$$

and this can be seen to be $\ll 1$ for any $m \geq 2$.

Now we want $c$ to satisfy the inequality $c^{-\beta} e^{\alpha c} \geq e^\gamma$, equivalently $f(c) \geq 0$. If $\alpha \geq \gamma$, $f(1) \geq 0$. Otherwise, we note that $f$ is strictly convex over $(0, \infty)$, has a minimum at $\beta/\alpha$, and increases thereafter. So if $f(\beta/\alpha) \geq 0$, $c_3 = \max(1, \beta/\alpha)$. If not, we have $f(1) < 0, f(\beta/\alpha) < 0$, so $f$ has a unique root in $(\max(1, \beta/\alpha), \infty)$ as indicated. [This root can be expressed in terms of the Lambert W-function, but the added complexity does not seem worth it.]

2. For (6.10), in §4.3.1 of [Oik17] and in the proof of eq. (4.26) there it is shown that the inequality $x \geq A \ln x + B$ is satisfied by

$$x = \begin{cases} 2A \ln(A+B) + B, & A+B \geq 1, \\ B + \ln B, & A+B < 1. \end{cases}$$

[The second line above is a slight improvement over the $1.5B + \ln B$ in [Oik17], obtained by noting that in the case here $B < 1$ so $\ln B < 0$.] (6.10) follows by taking $A = \beta/\alpha, B = \gamma/\alpha$.

## Appendix B   Proof of Lemma 5.2

Let $x^*$ maximize $G(x \| \mu)$ over a compact convex set $\mathcal{C}$. Assuming that $\mathcal{C}$ is contained in an open subset of $\mathbb{R}_+^m$, by a 2nd-order Taylor expansion around $x^*$ we have that for any $x \in \mathcal{C}, x \neq x^*$, there is a $\tilde{x} = ax + (1-a)x^*$ for some $a \in (0, 1)$ such that

$$G(x \| \mu) = G(x^* \| \mu) + \nabla_x G(x^* \| \mu) \cdot (x - x^*) + \frac{1}{2}(x - x^*)^T \nabla_x^2 G(\tilde{x} \| \mu)(x - x^*). \tag{B.1}$$

The term $\nabla_x G(x^*\|\mu) \cdot (x-x^*)$ is $\leqslant 0$ for any $x \in C$, as this is a necessary condition for $x^*$ to maximize $G(x\|\mu)$ over $C$. Further, the Hessian is

$$\nabla_x^2 G(\tilde{x}\|\mu) = \frac{1}{\sum_i \tilde{x}_i} \mathbb{1}_{m,m} - \text{diag}\left(\frac{1}{\tilde{x}_1}, ..., \frac{1}{\tilde{x}_m}\right),$$

where $\mathbb{1}_{m,m}$ is an $m \times m$ matrix of 1s. Therefore, writing $\tilde{x}$ as $x^* + a(x - x^*)$,

$$\forall x \in C, \quad G(x^*\|\mu) - G(x\|\mu) \geqslant -\frac{1}{2}(x-x^*)^T \nabla_x^2 G(\tilde{x}\|\mu)(x-x^*)$$
$$= \frac{1}{2}\left(\sum_i \frac{(x_i - x_i^*)^2}{x_i^* + a(x_i - x_i^*)} - \frac{(\sum_i x_i - \sum_i x_i^*)^2}{\sum_i x_i^* + a(\sum_i x_i - \sum_i x_i^*)}\right) \quad \text{(B.2)}$$
$$\triangleq \frac{1}{2} f(x;a).$$

[The notation $f(x;a)$ is somewhat misleading because once $x$ is given the possible $\tilde{x}$ and $a$ are determined, but it is intended to indicate that we treat $a$ as a parameter whose value is unknown.] We show in §B.4 that

$$\forall x \in \mathbb{R}_+^m, \forall a \in (0,1), \quad f(x;a) \geqslant 0 \quad \text{and} \quad f(x;a) = 0 \text{ iff } x = cx^* \text{ for some } c > 0. \quad \text{(B.3)}$$

[So the above Hessian, a function of $x, a$, is positive-semidefinite.] *Note* that (B.2) holds only for $x \in C$ whereas (B.3) holds for all $x \in \mathbb{R}_+^m$.

Using the notation $s \triangleq \sum_i x_i$, $s^* \triangleq \sum_i x_i^*$,

$$f(x;a) = \sum_i \frac{(x_i - x_i^*)^2}{x_i^* + a(x_i - x_i^*)} - \frac{(s - s^*)^2}{s^* + a(s - s^*)} \geqslant 0 \quad \forall a \in (0,1). \quad \text{(B.4)}$$

By Proposition 5.1(1) we want to minimize $f(x;a)$ over $C$ under the constraint $\|x - x^*\|_\infty = \vartheta \|x^*\|_\infty$. For the purposes of the proof we introduce an auxiliary constraint $s - s^* = t$, where we treat $t \in \mathbb{R}$ as a parameter. This parameter is constrained by $|t| \leqslant m\vartheta \|x^*\|_\infty$, since $|t| = \big|\|x\|_1 - \|x^*\|_1\big| \leqslant \|x - x^*\|_1 \leqslant m\|x - x^*\|_\infty$ [and also by (2.10)].

From this point on in the proof, *we relax the problem by ignoring the constraint $x \in C$*, except in a few instances clearly identified in the sequel. Then setting $u \triangleq x - x^*$, the minimum of $f(x;a,t)$ will be $\geqslant$

$$\min_{\|u\|_\infty = \vartheta\|x^*\|_\infty, \sum_i u_i = t, u > -x^*} \varphi(u) - \frac{t^2}{s^* + at}, \quad \text{where} \quad \varphi(u) \triangleq \sum_i \frac{u_i^2}{au_i + x_i^*} = \sum_i \varphi_i(u_i), \quad \text{(B.5)}$$

and we note that the minimum is over a non-convex set[B.1]. Now define the convex sets

$$\begin{aligned} C_j^+ &\triangleq \{u_j = \vartheta\|x^*\|_\infty, \sum_i u_i = t, u_i > -x_i^*, |u_i| \leqslant \vartheta\|x^*\|_\infty\} \\ C_j^- &\triangleq \{u_j = -\vartheta\|x^*\|_\infty, \sum_i u_i = t, u_i > -x_i^*, |u_i| \leqslant \vartheta\|x^*\|_\infty\} \end{aligned}, \quad j = 1, ..., m. \quad \text{(B.6)}$$

Then $u$ is in the non-convex set $\{\|u\|_\infty = \vartheta\|x^*\|_\infty, \sum_i u_i = t, u > -x^*\}$ of (B.5) iff it is in one of the $2m$ convex sets (B.6) [which are not disjoint].

We can find $\hat{u} = \text{argmin } \varphi(u)$ by minimizing $\varphi(u)$ over each one of the sets (B.6) and taking the minimum of the results. To do this, we will minimize $\varphi(u)$ over $C_1^+$ in §B.1 and over $C_1^-$ in §B.2, and then in §B.3 show how to generalize to the rest of the sets in (B.6).

---

B.1. Nevertheless $\varphi_i(u_i)$ is strictly convex for $u_i > -x_i^*$, and so $\varphi(u)$ is also strictly convex for $u > -x^*$.

## B.1 The case $u \in C_1^+$

To ease the notation we use $x_{\max}^* \triangleq \|x^*\|_\infty$ and $v \triangleq \vartheta \|x^*\|_\infty$. Then $C_1^+ = \{u_1 = v, \sum_i u_i = t, u_i > -x_i^*, |u_i| \leq v\}$, and with $\hat{u}_1 = v$ we have

$$\min_{u \in C_1^+} \varphi(u) = \varphi_1(v) + \min_{u \in C_1^+} \sum_{i \geq 2} \varphi_i(u_i). \tag{B.7}$$

For the second term we *relax the constraints* $|u_i| \leq v$, and introduce a Lagrange multiplier $\lambda$ for $\sum_{i \geq 2} u_i = t - v$. Then we must have

$$\frac{(a u_i + 2 x_i^*) u_i}{(a u_i + x_i^*)^2} + \lambda = 0, \quad i \geq 2,$$

which implies that $a(\lambda a + 1) u_i^2 + 2 x_i^* (\lambda a + 1) u_i + \lambda (x_i^*)^2 = 0$. If $\lambda a + 1 = 0$, this reduces to $\lambda (x_i^*)^2 = 0$, so we must have $\lambda = 0$; this leads either to $\hat{u}_i$ that violate $u_i > -x_i^*$, or to $\hat{u}_i = 0$ for $i \geq 2$, possible only if $t = v$. If $\lambda a + 1 < 0$ there is no real root, so it must be that

$$\hat{u}_i = \frac{\pm 1 - \sqrt{\lambda a + 1}}{a \sqrt{\lambda a + 1}} x_i^*, \quad \lambda \neq 0, \lambda a + 1 > 0, \quad i \geq 2.$$

The $\hat{u}_i$ with the $-1$ in the numerator does not satisfy $\hat{u}_i > -x_i^*$, so we are left with $\hat{u}_i = \frac{1 - \sqrt{\lambda a + 1}}{a \sqrt{\lambda a + 1}} x_i^*$, and using the constraint $\sum_{i \geq 2} u_i = t - v$ we find

$$\hat{u}_1 = v = \vartheta x_{\max}^*, \qquad \hat{u}_j = \frac{t - v}{s^* - x_1^*} x_j^*, \quad j \geq 2 \tag{B.8}$$

This includes the solution $\hat{u}_j = 0$ for $j \geq 2$ when $t = v$ found above. To satisfy $\hat{u}_i > -x_i^*$ we must have $t - v > x_1^* - s^*$; this reduces to $\hat{s} - x_1^* > v$ [B.2], which holds by virtue of $\hat{x}_1 - x_1^* = \hat{u}_1 = v$. Finally, from (B.8) and $\hat{u} = \hat{x} - x^*$,

$$\hat{x}_1 = x_1^* + v, \qquad \hat{x}_j = \left(1 + \frac{t - v}{s^* - x_1^*}\right) x_j^*, \quad j \geq 2. \tag{B.9}$$

We see that unless $t < v$, the point $\hat{x}$ is not admissible: by Proposition 2.2(4) we would have $G(\hat{x} \| \mu) > G(x^* \| \mu)$, so $\hat{x}$ cannot be in $C$. Therefore *we assume $t < v$ in the rest of this section*.

Now we return for a moment to the constraint $|u_j| \leq v$ for $j \geq 2$. By (B.8) it is equivalent to

$$\forall j \geq 2, \quad \left(1 - \frac{s^* - x_1^*}{x_j^*}\right) v \leq t \leq \left(1 + \frac{s^* - x_1^*}{x_j^*}\right) v,$$

so in case $t$ satisfies these conditions, the solution (B.8), (B.9) of the relaxed problem also solves the original problem (B.7).

With (B.8) the min in the second term of (B.7) reduces to $\frac{(t-v)^2}{s^* - x_1^* + a(t-v)}$ and so

$$\min_{u \in C_1^+} \varphi(u) \geq \frac{v^2}{x_1^* + a v} + \frac{(t - v)^2}{s^* - x_1^* + a(t - v)} > 0. \tag{B.10}$$

From (B.10) and (B.5),

$$f(x; a, t) \geq \frac{v^2}{x_1^* + a v} + \frac{(t - v)^2}{s^* + a t - (x_1^* + a v)} - \frac{t^2}{s^* + a t} \triangleq \psi(a, t, v). \tag{B.11}$$

---

B.2. Here we are using the covert notation $\hat{u} = \hat{x} - x^*$ and $\hat{s} = \sum_i \hat{x}_i$.

### B.1.1 Minimization w.r.t. $a$

We begin by minimizing $\psi(a,t,v)$ w.r.t. $a$, since once $x$ is given in (B.2), $a$ is restricted.

First we investigate $\psi(0,t,v)$ and $\psi(1,t,v)$:

$$\psi(0,t,v) = \frac{(vs^* - tx_1^*)^2}{x_1^* s^* (s^* - x_1^*)}, \qquad \psi(1,t,v) = \frac{(vs^* - tx_1^*)^2}{(x_1^* + v)(s^* + t)(s^* - x_1^* - v + t)}.$$

Clearly $\psi(0,t,v) \geqslant 0$ and it is 0 at $t = vs^*/x_1^*$, but this is excluded by the assumption $t < v$ made after (B.9). Therefore $\forall t, \psi(0,t,v) > 0$. For $\psi(1,t,v)$,

$$s^* - x_1^* + t - v > 0 \quad \forall t \tag{B.12}$$

because $s^* - x_1^* + t - v = s - x_1^* - v > 0$ since $x_1 - x_1^* = v$ implies $s \geqslant x_1^* + v$. Then $t = vs^*/x_1^*$ is excluded as noted above, and therefore $\forall t, \psi(1,t,v) > 0$.

Next, it is too hard to deal with $\partial \psi / \partial a$ as a function of $a$ directly, so we use a transformation with the notation

$$\sigma_1 \triangleq x_1^* + av, \quad \sigma_2 \triangleq s^* + at, \quad \rho(a) \triangleq \sigma_1 / \sigma_2, \qquad \sigma_1, \sigma_2, \rho > 0.$$

Then we have[B.3]

$$\psi(a,t,v) = \frac{v^2}{\sigma_1} - \frac{t^2}{\sigma_2} + \frac{(t-v)^2}{\sigma_2 - \sigma_1}, \qquad \frac{\partial \psi}{\partial a} = \frac{t^3}{\sigma_2^2} - \frac{v^3}{\sigma_1^2} - \frac{(t-v)^3}{(\sigma_2 - \sigma_1)^2}.$$

Further,

$$\frac{x_1^*}{s^*} \leqslant \rho(a) \leqslant \frac{x_1^* + v}{s^* + t} < 1, \quad \forall a, t. \tag{B.13}$$

This follows from $\rho'(a) = \frac{vs^* - tx_1^*}{(s^* + at)^2}$. If $t > 0$, $vs^* - tx_1^* = 0$ is disallowed as above, and for $t < (s^*/x_1^*)v$, $\rho'(a) > 0$. If $t < 0$, $\rho'(a) > 0$ always. So the minimum of $\rho(a)$ is $\rho(0)$ and its maximum is $\rho(1)$.

Returning to $\partial \psi / \partial a$,

$$\frac{\partial \psi}{\partial a} = \frac{(t\rho - v)^2 (t\rho^2 + 2(v-t)\rho - v)}{\sigma_2^2 (\rho - 1)^2 \rho^2}.$$

So the sign of $\partial \psi / \partial a$ is determined by that of $\varphi(\rho,t) = t\rho^2 + 2(v-t)\rho - v$ in the numerator. Now we distinguish some cases within $t < v$[B.4]:

**Case 1.** $0 < t < v$ and $\rho \leqslant 1/2$.

Then $\varphi(\rho,t) = t\rho(\rho - 2) + (2\rho - 1)v < 0$, so $\partial \psi / \partial a < 0$ and $\min_a \psi(a,t,v) = \psi(1,t,v)$. From (B.13) the assumption $\rho \leqslant \frac{1}{2}$ will hold if $2(x_1^* + v) \leqslant s^* + t$. With $t \in (0,v)$ this will hold if

$$\vartheta \leqslant \frac{1}{2} \frac{s^*}{x_{\max}^*} - 1 \tag{B.14}$$

which appears in the statement of the lemma and imposes a restriction on $x^*$: $x_{\max}^* < s^*/2$ [B.5].

**Case 2.** $t < 0$.

We show that $\psi(a, -t, v) > \psi(a, t, v)$ for all $t > 0$, and the same goes for the minima, so we don't need to do anything more for the $t < 0$ case, the already-derived bounds hold a fortiori.

---

B.3. Most of the following calculations were done with the help of the computer algebra system MAXIMA, https://maxima.sourceforge.io.

B.4. Recall assumption after (B.9).

B.5. Noting that $\varphi(\rho,t)$ is an affine function of $t$ controlled by the values $\varphi(\rho,0)$ and $\varphi(\rho,v) < 0$, we see that if $\rho > 1/2$ we may have $\min_a \psi(a,t,v) = \psi(0,t,v)$; this may open a way toward relaxing $x_{\max}^* < s^*/2$.

With $t>0$, $\psi(a,-t,v)>\psi(a,t,v)$ is equivalent to

$$\frac{(t+v)^2}{s^*-at-(x_1^*+av)} - \frac{(t-v)^2}{s^*+at-(x_1^*+av)} > t^2\left(\frac{1}{s^*-at}-\frac{1}{s^*+at}\right)$$

$$\iff \frac{2vs^*-2vx_1^*-av^2+at^2}{(s^*-at-(x_1^*+av))(s^*+at-(x_1^*+av))} > \frac{at^2}{(s^*-at)(s^*+at)}$$

$$\iff \frac{2v(s^*-x_1^*)-av^2+at^2}{at^2} > \frac{(s^*-at-(x_1^*+av))(s^*+at-(x_1^*+av))}{(s^*-at)(s^*+at)}.$$

The r.h.s. of the last inequality is $<1$, so it suffices if the l.h.s. is $\geqslant 1$. This reduces to $2(s^*-x_1^*)\geqslant av$, which holds if $\vartheta\leqslant 2(s^*/x_{\max}^*-1)$, a requirement weaker than (B.14).

**Case 3.** $t=0$.

Here we have $\varphi(\rho,0)=(2\rho-1)v$, so $\partial\psi/\partial a\leqslant 0$ iff $\rho\leqslant \tfrac{1}{2}$, which will hold by (B.14), and then $\min_a \psi(a,0,v)=\psi(1,0,v)$ again. [If $\rho>\tfrac{1}{2}$, $\partial\psi/\partial a>0$ and $\min_a \psi(a,0,v)=\psi(0,0,v)$.]

### B.1.2 Minimization w.r.t. $t\neq 0$

The conclusion from §B.1.1 is that we have to minimize $\psi(1,t,v)$, where $0\leqslant t<v$. [We shouldn't do anything w.r.t. $x=x_1^*$ at this point, because $x_1^*$ can't depend on $t=s-s^*$.]

(B.8), (B.9) specify a family of solutions $\hat{x}=\hat{x}(t)$, parameterized by $t$. Setting a value for $t$, e.g. by minimizing over $t$, picks out one of these solutions. We've already handled $t=0$, so here we have $0<t<v$. From (B.11),

$$\frac{\partial\psi(1,t,v)}{\partial t} = -\frac{(vs^*-tx_1^*)\big((2s^*-x_1^*)t-(2x_1^*+v)s^*+2s^{*2}\big)}{(s^*+t)^2(s^*-x_1^*-v+t)^2}.$$

The first factor in the numerator is positive by the assumption $t<v$. The second factor is positive iff $t>\frac{2x_1^*+v-2s^*}{2s^*-x_1^*}s^*$. But this is true because $t\geqslant 0$ and the numerator of the r.h.s. is negative if $\vartheta\leqslant 2(s^*/x_{\max}^*-1)$, as at the end of Case 2 in §B.1.1.

Consequently $\psi(1,t,v)$ decreases with increasing $t\in(0,v)$ and therefore

$$\inf_{t\in(0,v)} \psi(1,t,v) = \psi(1,v,v) = \left(\frac{1}{x_1^*+v}-\frac{1}{s^*+v}\right)v^2.$$

## B.2 The case $u\in C_1^-$

Here we have the set $C_1^-=\{u_1=-v,\ \sum_i u_i=t,\ u_i>-x_i^*,\ |u_i|\leqslant v\}$. Proceeding as in §B.1 and omitting the details, (B.8) and (B.9) become

$$\hat{u}_1==-v=-\vartheta x_{\max}^*, \qquad \hat{u}_j=\frac{t+v}{s^*-x_1^*}x_j^*, \quad j\geqslant 2,$$

and

$$\hat{x}_1=x_1^*-v, \qquad \hat{x}_j=\left(1+\frac{t+v}{s^*-x_1^*}\right)x_j^*, \quad j\geqslant 2.$$

There is no easy constraint on $t$ here for $\hat{x}$ to be in $C$, so we assume $t\in[-mv,mv]$. Next, (B.10) becomes

$$\min_{u\in C_1^-}\varphi(u) \geqslant \frac{v^2}{x_1^*-av}+\frac{(t+v)^2}{s^*-x_1^*+a(t+v)} > 0,$$

and (B.11) becomes

$$f(x;a,t) \geqslant \frac{v^2}{x_1^* - av} + \frac{(t+v)^2}{s^* + at - (x_1^* - av)} - \frac{t^2}{s^* + at} \triangleq \omega(a,t,v). \tag{B.15}$$

From (B.11) we see that $\omega(a,t,v) = \psi(a,t,-v)$, not entirely a surprise. Now we show that

$$t \geqslant 0 \implies \omega(a,t,v) > \psi(a,t,v), \quad \omega(a,-t,v) \geqslant \psi(a,t,v) \tag{B.16}$$

These inequalities allow us to short-circuit the rest of the derivations, as the bounds we have derived in §B.1 for $\psi(a,t,v)$ when $t \geqslant 0$ then hold a fortiori for $\omega(a,t,v)$ whether $t \geqslant 0$ or $t < 0$.

To derive the first inequality in (B.16), from (B.11) and (B.15),

$$\omega(a,t,v) - \psi(a,t,v) = \frac{2v(av^2 s^{*2} + 2tx_1^{*2} s^* - 2av^2 x_1^* s^* - 2tx_1^{*3} + at^2 x_1^{*2})}{(x_1^* - av)(x_1^* + av)(s^* - x_1^* - av + at)(s^* - x_1^* + av + at)},$$

so the sign of the difference is that of the 2nd factor in the numerator, an affine function of $a$:

$$(v^2 s^{*2} - 2v^2 x_1^* s^* + t^2 x_1^{*2})a + 2tx_1^{*2}(s^* - x_1^*).$$

This is positive at $a=0$, and also positive at $a=1$ if $t \geqslant 0$, which establishes the inequality.

For the second inequality,

$$\omega(a,-t,v) - \psi(a,t,v) =$$
$$\frac{2a(vs^* - tx_1^*)^2 \left(tv(t-v)a^2 + vs^{*2} + 2(t-v)x_1^* s^* - tx_1^{*2}\right)}{(x_1^* - av)(x_1^* + av)(s^* - at)(s^* + at)(s^* - x_1^* - av + at)(s^* - x_1^* + av - at)}.$$

The inequality is satisfied with equality at $t = (s^*/x_1^*)v$. Otherwise, in the third factor of the numerator the coefficient of $a^2$ is minimized at $t = v/2$ and equals $-v^3/4$. So the third factor is $\geqslant$

$$-v^2/4 + vs^{*2} + 2(t-v)x_1^* s^* - tx_1^{*2} = (2s^* - x_1^*)x_1^* t + (s^{*2} - 2x_1^* s^* - v^2/4)v$$
$$\geqslant (s^{*2} - 2x_1^* s^* - v^2/4)v$$

and the last expression is $\geqslant 0$ if $\vartheta \leqslant 2\frac{s^*}{x_1^*}\sqrt{1 - 2x_1^*/s^*}$. This is a weaker condition on $\vartheta$ than (B.14), but still requires $s^*/x_{\max}^* > 2$.

This completes the proof of (B.16) and of the case $u \in C_1^-$.

## B.3 Generalization to $C_j^+, C_j^-$ for $j > 1$.

It is clear that if at the beginning of §B.1 we had chosen the set $C_j^+$, $j \neq 1$, in (B.6), we would have simply ended up with $x_j^*$ instead of $x_1^*$ in (B.11) and then in §B.1.1 and §B.1.2. The same goes for choosing $C_j^-$, $j \neq 1$, in §B.2: we would have $x_j^*$ instead of $x_1^*$ in (B.15). To remove this dependence on $j$ altogether, we minimize the bounds we have obtained one final time, w.r.t. $x_1^*$.

There are two constraints on $x_1^*$: $x_1^* < s^* - v$ by (B.12), and $x_1^* \leqslant x_{\max}^*$. (B.14) implies that $s^* - \vartheta x_{\max}^* > x_{\max}^*$, so only the constraint $x_1^* \leqslant x_{\max}^*$ matters.

For $t = 0$ and $t \in (0,v)$ respectively, we have the functions

$$\left(\frac{1}{s^* - x_1^* - v} + \frac{1}{x_1^* + v}\right)v^2 = \psi(1,0,v), \qquad \left(\frac{1}{x_1^* + v} - \frac{1}{s^* + v}\right)v^2 = \psi(1,v,v).$$

The minimum of $\psi(1,0,v)$ occurs at $x_1^* = s^*/2 - v$, but by (B.14) this is $\geq x_{max}^*$, so the function is minimized at $x_{max}^*$ and has the value $\frac{\vartheta^2}{1+\vartheta}\frac{s^*}{s^* - (1+\vartheta)x_{max}^*}x_{max}^*$. The function $\psi(1,v,v)$ has its minimum at the largest allowable $x_1^*$, i.e. $x_{max}^*$, and its value is $\frac{\vartheta^2}{1+\vartheta}\frac{s^* - x_{max}^*}{s^* + \vartheta x_{max}^*}x_{max}^*$. Expressing these values in terms of $x_{max}^*/s^*$,

$$\min_{x_1^*}\psi(1,0,v) = \frac{\vartheta^2}{1+\vartheta}\frac{1}{1-(1+\vartheta)x_{max}^*/s^*}x_{max}^*, \qquad \min_{x_1^*}\psi(1,v,v) = \frac{\vartheta^2}{1+\vartheta}\frac{1-x_{max}^*/s^*}{1+\vartheta x_{max}^*/s^*}x_{max}^*.$$

The first bound is smaller than the second, and is our final result in the statement of the lemma. It is not unreasonable that this is so: the $t=0$ problem is more tightly constrained around $x^*$, and so the minimum difference $G(x^* \| \mu) - G(x \| \mu)$ is smaller.

## B.4 Proof of (B.3)

This is an improvement of the proof in [Oik17], Proposition 2.2.

Set $u = x - x^*$ and $\tilde{x} = x^* + a(x - x^*)$ so that (B.2) becomes

$$\begin{aligned} f(x;a) &= \sum_i \frac{u_i^2}{\tilde{x}_i} - \frac{(\sum_i u_i)^2}{\sum_j \tilde{x}_j} = \frac{1}{\sum_j \tilde{x}_j}\left(\sum_i \frac{u_i^2}{\tilde{x}_i/\sum_j \tilde{x}_j} - \left(\sum_i u_i\right)^2\right) \\ &= \frac{1}{\sum_j \tilde{x}_j}\left(\sum_i \frac{u_i^2}{\zeta_i} - \left(\sum_i u_i\right)^2\right) \triangleq \frac{1}{\sum_j \tilde{x}_j}g(\zeta), \end{aligned}$$

where $\zeta_i = \tilde{x}_i/\sum_j \tilde{x}_j$. For fixed $u$, $g(\zeta)$ is a strictly convex function of $\zeta \in \mathbb{R}_+^m$ and has a unique minimizer $\hat{\zeta}_i = u_i/\sum_j u_j$, at which point its value is 0. This establishes $f(x;a) \geq 0$ for any $x, a$.

To show that $f(x;a) = 0$ iff $x = cx^*$ for some $c > 0$, when $g(\zeta) = 0$ we have

$$\zeta_i = \frac{u_i}{\sum_j u_j} \iff \frac{x_i^* + a(x_i - x_i^*)}{s^* + a(s - s^*)} = \frac{x_i - x_i^*}{s - s^*},$$

where $s \triangleq \sum_i x_i$, $s^* \triangleq \sum_i x_i^*$. Now given any $c > 0$, $x = cx^*$ implies the last equality above. Conversely, if that equality holds, it implies that $x_i = (s/s^*)x_i^*$, so $c = s/s^*$.